\documentclass[11pt]{article}
\usepackage{geometry}
\usepackage{mathptmx}       
\usepackage{helvet}         
\usepackage{courier}        
\usepackage{type1cm}        
\usepackage{amsmath,amssymb}
\usepackage{mathrsfs}
\usepackage{amsfonts}
\usepackage{makeidx}         
\usepackage{graphicx}        
\usepackage{multicol}        
\usepackage[bottom]{footmisc}
\usepackage{caption}
\usepackage{subcaption}
\usepackage{epsfig}
\usepackage{bm}
\usepackage{enumitem}
\usepackage{epstopdf}
\usepackage{multirow}
\usepackage{booktabs}
\usepackage{mathtools}
\usepackage{wrapfig}
\usepackage{titlesec}
\usepackage[round]{natbib}

\mathtoolsset{showonlyrefs}
\newtheorem{theorem}{Theorem}

\newtheorem{lemma}{Lemma}

\newtheorem{condition}{Condition}[section]

\titlespacing\section{0pt}{11pt plus 0pt minus 0pt}{0pt plus 2pt minus 2pt}
\titlespacing\subsection{0pt}{11pt plus 4pt minus 2pt}{0pt plus 2pt minus 2pt}
\titlespacing\subsubsection{0pt}{11pt plus 4pt minus 2pt}{0pt plus 2pt minus 2pt}

\titleformat{\section}{\normalfont\fontsize{12}{12}\bfseries}{\thesection}{1em}{}
\titleformat{\subsection}{\normalfont\fontsize{11}{11}\itshape}{\thesubsection}{1em}{}
\titleformat{\subsubsection}{\normalfont\fontsize{10}{10}\itshape}{\thesubsubsection}{1em}{}

\begin{document}

\small{

\begin{center}
\textbf{Conditional Information and Inference in Response-Adaptive Allocation Designs}
\end{center}

\begin{center}
{Adam Lane}\\
{Cincinnati Children's Hospital Medical Center}\\
{\today} \\
\vspace{0.5cm}
\end{center}

\begin{abstract}
Response-adaptive allocation designs refer to a class of designs where the probability an observation is assigned to a treatment is changed throughout an experiment based on the accrued responses. Such procedures result in random treatment sample sizes. Most of the current literature considers unconditional inference procedures in the analysis of response-adaptive allocation designs. The focus of this work is inference conditional on the observed treatment sample sizes. The inverse of information is a description of the large sample variance of the parameter estimates. A simple form for the conditional information relative to unconditional information is derived. It is found that conditional information can be greater than unconditional information. It is also shown that the variance of the conditional maximum likelihood estimate can be less than the variance of the unconditional maximum likelihood estimate. Finally, a conditional bootstrap procedure is developed that, in the majority of cases examined, resulted in narrower confidence intervals than relevant unconditional procedures.
\end{abstract}

\section{Introduction}

Response-adaptive allocation designs alter the probability an observation is assigned to the available treatments based on the accrued responses. Such designs have gained popularity in clinical trials due to their ability to balance ethical and experimental objectives; as a consequence observations will be referred to as subjects. This paper considers response-adaptive allocation designs with binary responses. In phase I clinical trials the toxicity profile of a drug is examined across multiple dose levels. In the context of this paper each dose level would be considered a different treatment. The phase I experimental objective is to determine the maximum tolerated dose (MTD) - defined as the highest dose with toxicity probability less than a predetermined percent. A common phase I ethical objective is to maximize the number of subjects treated at the MTD. In  phase II and III clinical trials the efficacy of a set of treatments is considered. The experimental objective is to estimate the treatment success probabilities. The ethical objective is to minimize the number of treatment failures. Ethical considerations can be incorporated into response-adaptive allocation designs and balanced against the more traditional experimental objectives. The randomized play the winner (RPW) rule, [\citet{Wei:Adap:1978}], Success Driven designs (SDD) [\citet{Durh:Flou:Li:Sequ:1998}], up and down designs [\citet{Durh:Flou:rand:1994}], sequential maximum likelihood procedures [\citet{Rose:Stal:Ivan:Opti:2001}] and doubly adaptive biased coin designs [\citet{Eise:TheD:1994} and \citet{Hu:Zhan:Asym:2004}], are a small collection of available binary response-adaptive allocation designs.

A response-adaptive allocation design is defined by a random or deterministic rule for allocating subjects to treatments. As an illustration consider the RPW($\alpha$,$\beta$) rule defined for treatments $1$ and $2$ with success probabilities $p_{1}$ and $p_{2}$, respectively. The RPW rule assigns treatments sequentially based on random draws from an adaptive urn. The initial urn is comprised of $\alpha$ balls of each type, $1$ and $2$, representing the corresponding treatment. When a response is observed $\beta$ type $1$ balls are added if either; a success is observed on treatment $1$; or a failure is observed on treatment $2$; otherwise $\beta$ type $2$ balls are added. Implementing a RPW rule in a clinical trial can significantly reduce the total number of expected treatment failures. Define $S_{k}(i)$ and $N_{k}(i)$ to be the number of successes and the total number of trials from the first $i$ subjects on treatment $k=(1,2)$, respectively. It is well known that the unconditional maximum likelihood estimate (MLE) is the same as if no adaptation had taken place, i.e., the MLE from a sample of size $n$ is $\hat{p}_{k}=S_{k}(n)/N_{k}(n)$.

Using an adaptive design increases the complexity of inference since responses are no longer independent and treatment sample sizes are random. Not only are the sample sizes random, but they are not ancillary, since their distribution depends on the underlying success probabilities. There are two approaches to inference, conditional and unconditional. In the conditional approach inference is conducted with respect to a reference set that contains only the experiments with sample sizes equal to the observed sample sizes. The unconditional approach considers the set all possible experiments regardless of the sample sizes. In the context of response-adaptive allocation designs \citet{Wei:Smyt:Lin:Stat:1990}, \citet{Begg:OnIn:1990}, \citet{Flem:OnIn:1990}, \citet{Rose:Lach:Rand:2002}, \citet{Anto:Giov:OnTh:2006} and \citet{Pros:Dodd:ReRa:2019} argue that the non ancillary nature of the sample sizes implies unconditional inference is preferred to conditional inference. This paper finds that their intuitive argument does not hold in general.  Specifically, it is shown that conditional information can be greater than unconditional information; the variance of the conditional MLE can be less than the variance of the unconditional MLE; and conditional confidence intervals can be narrower than relevant unconditional intervals.

Conditional estimation and inference has received significant attention in adaptive designs with interim analyses. \citet{Cohe:Sack:TwoS:1989}, \citet{Stri:Case:Cond:2003},  \citet{Pepe:Feng:Long:Cond:2009}, \citet{Kima:Todd:Stal:Cond:2013}, \citet{Robe:Prev:Bowd:Acco:2016} and \citet{Mars:Scho:Unde:2018} propose conditional estimates and inference procedures in the context of group sequential designs. Group sequential designs are characterized by a series of interim analysis where the study terminates and/or the less successful treatment(s) are dropped from subsequent stages based on the observed success probabilities from the preceding stages. \citet{Scho:Mars:Meta:2013} examine bias in meta-analysis of group sequential designs with early stopping rules. \citet{Koop:Feng:Pepe:Cond:2012} and \citet{Brob:Mill:Cond:2017} propose conditional methods in sample size re-estimation designs. Sample size re-estimation is an adaptive design schema where the sample size for the second stage is calculated based on the success probabilities observed in the initial stage. Group sequential and sample size re-estimation designs will be explicitly excluded by the conditions, stated in Section \ref{sec:Pre}, required here. However, these works serve as motivation for examining conditional inference in response-adaptive allocation designs.

In response-adaptive allocation designs conditional inference has been examined in the context of the RPW rule. \citet{Wei:Smyt:Lin:Stat:1990}, heuristically, found exact unconditional inference to be superior to conditional inference. However, in Sections \ref{sec:Info} and \ref{sec:sim} it is shown that had \citet{Wei:Smyt:Lin:Stat:1990} considered a more exhaustive set of treatment success probabilities they would have found cases where conditional inference was superior. In contrast to \citet{Wei:Smyt:Lin:Stat:1990}, \citet{Begg:OnIn:1990} presents some of the negative aspects of unconditional inference for the RPW rule.  Briefly, \citet{Begg:OnIn:1990} argued that it is difficult to properly order the sample sample. Specifically, it is not sufficient to compare the observed $\hat{p}_{k}$ to $\hat{p}_{k}$ from an unobserved point in the sample space to reach a conclusion regarding the relative evidence in favor or against a specific hypothesis. This problem with ordering can result in un-appealing inference properties. This ordering problem does not occur on the conditional space.

Section \ref{sec:Pre} reviews binary response-adaptive allocation designs. In Section \ref{sec:Info} it is found that the large sample information loss (or gain) in the conditional distribution depends on the observed sample sizes. For cases where there is conditional information gain the conditional MLE is presented as alternative to the unconditional MLE. In Section \ref{sec:Est} a conditional bootstrap method is developed that generates consistent confidence intervals for the conditional MLE. In Section \ref{sec:sim} the distribution of the conditional MLE relative to the unconditional MLE is examined for three well known response adaptive allocation designs. For each design considered, there exist cases where the conditional MLE has a smaller variance than the unconditional MLE. Further, in the majority of cases considered the proposed conditional bootstrap results in narrower confidence intervals than relevant unconditional alternatives. Additionally, the conditional MLE was observed to have uniformly smaller bias. Finally, in Section \ref{sec:real} the fluoxetine placebo controlled clinical trial [\citet{Tamu:Fari:Ande:Heil:ACas:1994}] is used to demonstrate the benefits of conditional inference for a real world data set.

\section{Preliminaries} \label{sec:Pre}
In this section  the general underlying conditions and notation for the class of response-adaptive allocation designs considered in this paper are described. A sample of $n$ subjects will be assigned to one of $K$ treatments, indexed by $1,\ldots,K$. The treatment assignments occur sequentially, meaning some of the responses and treatments of the first $i-1$ subjects are known and used in the allocation of the $i$th subject to treatment. The distinction that only some of the responses might be available is made in order to include delayed response and batch designs.

The response of the $i$th the subject is denoted $Y_{i} (=0,1)$. The treatment received by the $i$th subject is described by a vector $\boldsymbol{T}(i) = (T_{i1},\ldots,T_{iK})$, where $T_{ik}=1$ if treatment $k$ is received by subject $i$ and 0 otherwise. Exactly one of $T_{ik}$, $k=1,\ldots,K$ is equal to one.

Further, denote the total number of success on treatment $k$ from the first $i$ subjects as $S_{k}(i) = \sum_{j=1}^{i} Y_{j}T_{jk}$; let $\boldsymbol{S}(i) = \{S_{1}(i),\ldots,S_{K}(i)\}^{T}$. The corresponding number of the first $i$ subjects receiving treatment $k$ is denoted $N_{k}(i) = \sum_{j=1}^{i} T_{jk}$ and $\boldsymbol{N}(i) = \{N_{1}(i),\ldots,N_{K}(i)\}^{T}$. Let $\boldsymbol{X}(n) = \{S_{1}(n),\ldots,S_{K}(n),N_{1}(n),\ldots,N_{K-1}(n)\}$ and $\mathscr{F}(i) = \sigma\{\boldsymbol{X}(1),\ldots,\boldsymbol{X}(i)\}$. Throughout $\boldsymbol{S}(n)$ and $\boldsymbol{N}(n)$ represents random variables and $\boldsymbol{s}$ and $\boldsymbol{n}$ denote their corresponding observed values. At times the dependence on $(n)$ is dropped from $\boldsymbol{S}(n)$ and $\boldsymbol{N}(n)$ when the meaning is clear.

Formally, a response-adaptive allocation design is given by the allocation rule that sequentially assigns subjects to treatment, i.e., a specific response-adaptive allocation design is defined by
\begin{align}
P\{T_{ik} = 1 | \mathscr{F}(i-1) \}
\end{align}
for $i=1,\ldots,n$ and $k=1,\ldots,K$. Note this probability can have support [0,1]. The case of 0,1 is a deterministic design. If it is (0,1) it is a randomized design. Further, for shorthand let $P_{\boldsymbol{p}}$ denote the probability measure of treatment responses and $E_{\boldsymbol{p}}$ denote the expectation with respect to $P_{\boldsymbol{p}}$. In this work the following conditions are required throughout.
\begin{condition}\label{C1}\mbox{}
\begin{enumerate}[nolistsep]
\item The total sample size, $n$, is fixed. \label{C1.0}
\item The number of treatments, $K$, is finite. \label{C1.2}
\item The conditional responses $Y_{i}|\{T_{i-1,k} = 1\}$ are independent with $P\{Y_{i}=1|T_{i-1,k} = 1\} = p_{k}\in (0,1)$ for all $i=1,\ldots,n$ and $k=1,\ldots,K$. \label{C1.3}
\item Conditional on the observed data from the first $i-1$ subjects the distribution of $P\{T_{ik} = 1 | \mathscr{F}(i-1) \}$ does not depend on $\boldsymbol{p} = (p_{1},\ldots,p_{K})$ \label{C1.4}.
\item $E_{\boldsymbol{p}}$ is continuous in $\boldsymbol{p}$.
\end{enumerate}
\end{condition}
These conditions are not overly restrictive; however, some designs are excluded. The first condition excludes group sequential and sample size re-estimation designs. The second condition excludes dose finding phase I or phase II trials where treatments are selected from a continuous dose space. The third excludes designs with non-binary responses. The class of designs meeting these criteria is broad and includes all of the designs mentioned in the introduction as well as many others.

It is well documented that the joint likelihood from an adaptive experiment is proportional to the likelihood of an experiment that contained no adaptation [see \citet{Ford:Titt:Wu:infe:1985}]. Therefore, under conditions \ref{C1}, the joint likelihood, from $n$ subjects, is
\begin{align}
L_{f}\{\boldsymbol{p};\boldsymbol{s},\boldsymbol{n}\} \propto \prod_{k=1}^{K} p_{k}^{s_{k}}(1-p_{k})^{n_{k} - s_{k}}.
\end{align}
However, only the likelihoods are the same; in response-adaptive allocation designs the minimal sufficient statistic for $\bm{p}$ is $\boldsymbol{X}(n)$. The distribution of $\boldsymbol{X}(n)$ differs significantly from a design with $n$ independently distributed Bernoulli trials.

The score function, the derivative of the log-likelihood, for adaptive designs satisfying conditions \ref{C1}, is
\begin{align}
\boldsymbol{\Psi}(\boldsymbol{\hat{p}},\boldsymbol{n}) = \frac{\partial}{\partial \boldsymbol{p}} \log L_{f}(\boldsymbol{p};\boldsymbol{s},\boldsymbol{n}) = \Lambda_{\boldsymbol{n}} (\boldsymbol{\hat{p}} - \boldsymbol{p}),
\end{align}
where $\boldsymbol{\hat{p}} = (\hat{p}_{1},\ldots,\hat{p}_{K})^{T}$, $\hat{p}_{k} = s_{k}/n_{k}$ and $\Lambda_{\boldsymbol{n}}$ is a $K\times K$ diagonal matrix with $k$th diagonal entry $n_{k}/[p_{k}(1-p_{k})]$. Setting $\boldsymbol{\Psi}=0$ demonstrates that $\boldsymbol{\hat{p}}$ is the unconditional maximum likelihood estimate (UMLE) of $\boldsymbol{p}$. However, $\boldsymbol{\hat{p}}$ is not one to one with the minimal sufficient statistic $\boldsymbol{X}(n)$ and is therefore not sufficient.

\subsection{Illustrative Examples}

The first example is the RPW rule described in the introduction. For the RPW rule the $i$th subject is allocated to treatment $1$ with probability
\begin{equation}
P\{T_{i1} = 1 | \mathscr{F}(i-1) \} = \frac{\alpha + \beta [S_{1}(i-1) + N_{2}(i-1) - S_{2}(i-1) ]}{ 2\alpha + i\beta },
\end{equation}

The second example is the SDD. The SDD is an adaptive urn design similar to the RPW rule. The motivation behind this design is that a failure on one treatment does not provide information about the probability of a success for any other treatment. A SDD($\alpha$,$\beta$), defined for two treatments, has an initial urn comprised of $\alpha$ balls of each type, $1$ and $2$, representing the corresponding treatment. If a success is observed on treatment 1 (2) then $\beta$ type 1 (2) balls are added otherwise the urn remains unchanged. For the SDD design the $i$th subject is allocated to treatment $k$ with probability
\begin{equation}
P\{T_{i1} = 1 | \mathscr{F}(i-1) \} = \frac{\alpha + \beta S_{1}(i-1)}{ 2\alpha + \beta \{S_{1}(i-1) + S_{2}(i-1)\} }.
\end{equation}

The third design is a from the class of sequential maximum likelihood procedures (SMLP) targeting efficiency. As stated by \citet{Melf:Page:1998}, for independent binary responses the variance of the $\hat{p}_{2} - \hat{p}_{1}$ is minimized when
\begin{align}
r_{1} = \frac{\sqrt{p_{1}(1-p_{1})}}{\sqrt{p_{1}(1-p_{1})} + \sqrt{p_{2}(1-p_{2})}}
\end{align}
proportion of subjects are assigned to treatment 1 and $1-r_{1}$ proportion are assigned to treatment 2. This allocation scheme is referred to as \textit{Neyman allocation}. In practice it is not possible to assign subjects according to this allocation since $p_{1}$ and $p_{2}$ are unknown. Instead, let $\hat{p}_{k}'(i) = [S_{k}(i) + 1/2] / [N_{k}(i) + 1]$ and define the probability the $i$th patient is allocated to treatment 1 as
\begin{align}
P\{T_{i1} = 1 | \mathscr{F}(i-1) \} = \hat{r}_{1}(i) = \frac{\sqrt{\hat{p}_{1}'(i-1)[1-\hat{p}_{1}'(i-1)]}}{\sqrt{\hat{p}_{1}'(i-1)[1-\hat{p}_{1}'(i-1)]} + \sqrt{\hat{p}_{2}'(i-1)[1-\hat{p}_{2}'(i-1)]}}.
\end{align}
This will be referred to as the Neyman allocation design (NAD).

\section{Information} \label{sec:Info}
This section presents unconditional and conditional information in response adaptive allocation designs. There are two common measures of information; observed and expected, both measures will be introduced and described. The main result of this section is a description of the asymptotic efficiency of the conditional information relative to the unconditional information.

\subsection{Unconditional Information}
Observed Fisher information is defined as the second derivative of the log-likelihood evaluated at the UMLE. Under the current framework it is straightforward to show that observed Fisher information is
\begin{align} \label{eq:ObsInfo}
J_{\boldsymbol{S},\boldsymbol{N}}(\boldsymbol{\hat{p}}) &= \left. \frac{\partial}{\partial \boldsymbol{p}} \boldsymbol{\Psi}(\boldsymbol{p},\boldsymbol{n}) \right|_{\boldsymbol{p}=\boldsymbol{\hat{p}}} \\ &= \hat{\Lambda}_{\boldsymbol{n}},
\end{align}
where $\hat{\Lambda}_{\boldsymbol{n}}$ is a $K\times K$ diagonal matrix with $k$th diagonal entry $n_{k}/[\hat{p}_{k}(1-\hat{p}_{k})]$. We will refer to this as \textit{unconditional observed information}.

In general, the total expected Fisher information of the joint distribution is by definition the variance of the score function, i.e.,
\begin{align} \label{eq:ExpInfo}
I_{\boldsymbol{S},\boldsymbol{N}}(\boldsymbol{p}) = \mbox{Var}[\boldsymbol{\Psi}].
\end{align}
Under conditions \ref{C1} the derivatives and integrals can be exchanged in \eqref{eq:ExpInfo} and as shown in \citet{Anto:Giov:OnTh:2006} it can be rewritten as
\begin{align} \label{eq:ExpInfoRed}
I_{\boldsymbol{S},\boldsymbol{N}}(\boldsymbol{p}) = \mbox{E}[ \Lambda_{\boldsymbol{N}} ].
\end{align}
This will be referred to as the \textit{unconditional expected information}. Practically, finding this expectation can be technically challenging. For this reason the unconditional observed information is commonly used in practice. Let $\nu_{k} = \lim_{n\rightarrow\infty} n^{-1}E\left[N_{k}(n)\right]$, where $\nu_{k}\in[0,1]$. Unconditional observed and expected information are asymptotically equivalent when the following conditions hold in addition to \ref{C1}.
\begin{condition}\label{C2}\mbox{}
\begin{enumerate}[nolistsep]
\item $N_{k}(n)/n \rightarrow \nu_{k}, k=1,\ldots,K$ in probability as $n\rightarrow\infty$. \label{C2.1}
\item $\boldsymbol{\hat{p}} \rightarrow \boldsymbol{p}$ in probability as $n\rightarrow\infty$. \label{C2.2}
\end{enumerate}
\end{condition}
\citet{Athr:Karl:Embe:1968} and \citet{Melf:Page;Gera:AnAd:2001} can be used to determine when 3.1.1 holds almost surely. \citet{Melf:Page:Esti:2000} provide powerful results to determine if 3.1.2 holds, again almost surely. For many response adaptive allocation designs condition \ref{C2} holds and the unconditional expected and observed information are asymptotically equivalent.

Both measures of information can be used to find the asymptotic distribution of the UMLE. Let
\begin{align}
\boldsymbol{Z} &= \Lambda_{\boldsymbol{N}}^{1/2}(\boldsymbol{\hat{p}} - \boldsymbol{p}) \\
\boldsymbol{\hat{Z}}_{exp} &= \{I_{\boldsymbol{S},\boldsymbol{N}}(\boldsymbol{\hat{p}})\}^{1/2}(\boldsymbol{\hat{p}} - \boldsymbol{p}) \\
\boldsymbol{\hat{Z}}_{obs} &= \{J_{\boldsymbol{S},\boldsymbol{N}}(\boldsymbol{\hat{p}})\}^{1/2}(\boldsymbol{\hat{p}} - \boldsymbol{p}).
\end{align}
\citet{Rose:Flou:Durh:Asym:1997} provide conditions that ensure that
\begin{align} \label{eq:asyEFI}
\boldsymbol{Z}_{j} \rightarrow N(\boldsymbol{0},\boldsymbol{I_{K}}),
\end{align}
where  $\boldsymbol{I}_{K}$ is the $K\times K$ identity matrix and $\boldsymbol{Z}_{j}$ can be $\boldsymbol{Z}$, $\boldsymbol{\hat{Z}}_{exp}$ or $\boldsymbol{\hat{Z}}_{obs}$. This justifies the use unconditional observed and expected information for for hypothesis testing and inference for many large sample response-adaptive allocation designs.

\citet{May:Flou:Asym:2009} show that in certain cases when the \citet{Rose:Flou:Durh:Asym:1997} conditions are not satisfied that \eqref{eq:asyEFI} still holds for $\boldsymbol{\hat{Z}}_{obs}$. In adaptive designs with normal errors \citet{LinFlouRose2019} also found that normalizing by the unconditional observed information results in asymptotic normality. This indicates that unconditional observed information may have broader application in the context of response-adaptive allocation designs than the unconditional expected information.

\subsection{Conditional Information}
In this work, conditional information is defined as the information in the conditional likelihood
\begin{align} \label{eq:condLik}
L_{c}(\boldsymbol{p};\boldsymbol{s},\boldsymbol{n}) = P\{\boldsymbol{S}(n) = \boldsymbol{s} | \boldsymbol{N}(n) = \boldsymbol{n}\}.
\end{align}
Conditional expected and observed information, with respect to $L_{c}$, have not been well described, in general, either exactly or approximately. In this section a convenient expression for the conditional expected and observed information is derived.

Denote the conditional score function as
\begin{align}
\boldsymbol{\Gamma}(\boldsymbol{p},\boldsymbol{s},\boldsymbol{n}) &= \frac{\partial}{\partial \boldsymbol{p}} \log L_{c}(\boldsymbol{p};\boldsymbol{s},\boldsymbol{n}).
\end{align}
By definition the conditional expected and observed information are
\begin{align}
I_{\boldsymbol{S}|\boldsymbol{N}}(\boldsymbol{\hat{p}}) = \mbox{Var}[\boldsymbol{\Gamma}]
\end{align}
and
\begin{align}
J_{\boldsymbol{S},\boldsymbol{N}}(\boldsymbol{\hat{p}}) &= \left. \frac{\partial}{\partial \boldsymbol{p}} \boldsymbol{\Gamma}(\boldsymbol{p},\boldsymbol{s},\boldsymbol{n}) \right|_{\boldsymbol{p}=\boldsymbol{\hat{p}}},
\end{align}
respectively. However, the above definitions are not easily calculated and do not provide immediate insight into the conditional information relative to their unconditional counterpart. The following theorem presents useful forms for the conditional observed and expected information.
\begin{theorem} \label{thm:CondInfo} Under Condition \ref{C1} and if $0<n_{k}<n$ or all $k=1,\ldots,K$ then
\begin{align} \label{eq:condExpInfo}
I_{\boldsymbol{S}|\boldsymbol{N}}(\boldsymbol{p}) &= \Lambda_{\boldsymbol{n}} {\rm{Var}}[\boldsymbol{\hat{p}}|\boldsymbol{N}(n)] \Lambda_{\boldsymbol{n}}
\end{align}
and
\begin{align} \label{eq:condObsInfo}
J_{\boldsymbol{S}|\boldsymbol{N}}(\boldsymbol{\hat{p}}) = I_{\boldsymbol{S}|\boldsymbol{N}}(\boldsymbol{\hat{p}}) - J_{\boldsymbol{S},\boldsymbol{N}}(\boldsymbol{\hat{p}}) D(\boldsymbol{\hat{p}}) .
\end{align}
where $D(\boldsymbol{p}) = \mbox{E}_{\boldsymbol{p}}\left[ (\boldsymbol{\hat{p}} - \boldsymbol{p}) | \boldsymbol{N}(n)=\boldsymbol{n} \right][(\partial/\partial \boldsymbol{p}) \Lambda_{\boldsymbol{n}}$.
\end{theorem}
This theorem provides expressions for the conditional information measures that do not require differentiation. Using these expressions the computation time required to compute $I_{\boldsymbol{S}|\boldsymbol{N}}(\boldsymbol{p})$ and $J_{\boldsymbol{S}|\boldsymbol{N}}(\boldsymbol{\hat{p}})$ are significantly reduced. Monte Carlo methods can be used to compute these quantities approximately which can further reduces computation time.

Theorem \ref{thm:CondInfo} shows that the difference between conditional observed information and conditional expected information is proportional to the conditional bias of the UMLE. If the UMLE has significant bias then the two can differ significantly. Under condition \ref{C2} the conditional bias will converge to 0 as $n\rightarrow\infty$; therefore for large $n$ the two conditional information measures will be equal for many designs.


More than one conditional paradigm exists. \citet{Silv:Opti:1980} and \citet{Ford:Titt:Wu:infe:1985} considered the appropriateness of ``ignoring'' the adaptive nature of the design. This can be formalized as conditional information by considering the conditional information of the $i$th subject. First, the conditional likelihood for $i$th subject is proportional to
\begin{align}
L_{i}(\boldsymbol{p};y_{i},\boldsymbol{t_{i}}) \propto \prod_{k=1}^{K} p_{k}^{y_{i}t_{ik}}(1-p_{k})^{1 - y_{i}t_{ik}}.
\end{align}
Then the expected information from subject $i$ conditioned on $\boldsymbol{t_{i}}$  is
\begin{align}
\boldsymbol{m_{i}}(p_{i};y_{i},\boldsymbol{t_{i}}) &= E\left[ \left. \left\{\frac{d}{d\boldsymbol{p}}\log L_{i}(p_{i};Y_{i},\boldsymbol{T_{i}}) \right\}\left\{\frac{d}{d\boldsymbol{p}}\log L_{i}(p_{i};Y_{i},\boldsymbol{T_{i}}) \right\}^{T}  \right| \boldsymbol{T_{i}} = \boldsymbol{t_{i}}  \right] \\
&= \Lambda_{\boldsymbol{t_{i}}},
\end{align}
where $\Lambda_{\boldsymbol{t_{i}}}$ is a diagonal matrix with $t_{ik}/[\hat{p}_{k}(1-\hat{p}_{k})]$ on the $k$th diagonal. Recall, only one $t_{ik}$, $k=1,\ldots,K$ can be equal to one. Therefore, the total per subject conditional information is
\begin{align}
M(\boldsymbol{p};\boldsymbol{y},\boldsymbol{t}) = \sum_{i=1}^{n} m_{i}(p_{i};y_{i},t_{ik}) = \Lambda_{\boldsymbol{n}}.
\end{align}
This is exactly the same as the expected information from a design with $\boldsymbol{n}$ fixed in advance. In the preceding section it was shown that for many designs  $\Lambda_{\boldsymbol{n}}$ is asymptotically equivalent to both $I_{\boldsymbol{S},\boldsymbol{N}}(\boldsymbol{p})$ and $J_{\boldsymbol{S},\boldsymbol{N}}(\boldsymbol{\hat{p}})$. This form of conditional expected information has been considered in \citet{Anto:Giov:OnTh:2006} where they have shown, in a more general setting, that asymptotically unconditional expected information and expected information conditioned on the set of observed treatment allocations are asymptotically equivalent.

The next section demonstrates that the asymptotic behavior of $I_{\boldsymbol{S}|\boldsymbol{N}}(\boldsymbol{p})$ and $J_{\boldsymbol{S}|\boldsymbol{N}}(\boldsymbol{\hat{p}})$ are not asymptotically equivalent to $I_{\boldsymbol{S},\boldsymbol{N}}(\boldsymbol{p})$ and $I_{\boldsymbol{S},\boldsymbol{N}}(\boldsymbol{\hat{p}})$. This distinguishes the conditional inference in this paper from the information obtained by ignoring the adaptive nature of the design.

\subsection{Relative Information}

As previously remarked, \citet{Wei:Smyt:Lin:Stat:1990} and \citet{Begg:OnIn:1990} found exact conditional inference for the RPW rule to be inferior to exact unconditional inference. \citet{Wei:Smyt:Lin:Stat:1990}, \citet{Flem:OnIn:1990}, \citet{Rose:Lach:Rand:2002} and \citet{Anto:Giov:OnTh:2006} have suggested there is significant information loss in the conditional distribution. Such comments are intuitive; however, it is not immediately clear if they are rigourous. In this section the previously derived expressions of information are used to define a measure of relative efficiency. The proposed definition of the relative efficiency will be used to examine these statements regarding conditional information more explicitly.

As previously suggested, under conditions \ref{C1} and \ref{C2} $\Lambda_{\boldsymbol{N}}$, $I_{\boldsymbol{S},\boldsymbol{N}}(\boldsymbol{p})$ and $J_{S,N}(\boldsymbol{\hat{p}})$, scaled by $n^{-1}$ converge to the same constant matrix. Under the same conditions $I_{\boldsymbol{S}|\boldsymbol{N}}(\boldsymbol{p})$ and $J_{S|N}(\boldsymbol{\hat{p}})$, scaled by $n^{-1}$, are asymptotically equivalent. However, it is not clear if the conditional measures converge to a constant matrix, which is a regularity condition traditionally required to establish the asymptotic normality of the conditional MLE [see \citet{Ande:Asym:1970}]. The relative efficiency defined in this section will also be used to investigate the convergence of the conditional information measures.

Under conditions \ref{C1} and \ref{C2} the \textit{expected relative efficiency}, defined as $I_{\boldsymbol{S}|\boldsymbol{N}}(\boldsymbol{p})\{I_{\boldsymbol{S},\boldsymbol{N}}(\boldsymbol{p})\}^{-1}$, and the \textit{observed relative efficiency}, defined as $J_{\boldsymbol{S}|\boldsymbol{N}}(\boldsymbol{\hat{p}})\{J_{\boldsymbol{S},\boldsymbol{N}}(\boldsymbol{\hat{p}})\}^{-1}$, are both asymptotically equivalent to
\begin{align} \label{eq:RelEff}
\mbox{Rel-Eff} = \mbox{Var}[\boldsymbol{Z}|\boldsymbol{N}(n)].
\end{align}
The proof is straightforward and omitted. We will refer to \eqref{eq:RelEff} as the \textit{relative efficiency} since it is asymptotically equivalent to both observed and expected relative efficiency. Note that the relative efficiency is a symmetric matrix.  When the relative efficiency is greater, say in Loewner order, than the identity matrix, i.e. $\mbox{Var}[\boldsymbol{Z}|\boldsymbol{N}(n)] \ge \boldsymbol{I}_{K}$, it indicates that conditional information is greater than unconditional. The converse is true if it is less than $\boldsymbol{I}_{K}$.

This definition of relative efficiency also provides insight into the difficulty in assessing asymptotic relative information gain/loss rigourously. As stated in equation (3), conditional observed and expected information, scaled by the corresponding unconditional information, are asymptotically equal to $\mbox{Var}[\boldsymbol{Z}|\boldsymbol{N}(n)]$. Even under the independent and identically distributed (i.i.d.) assumption there are few results in the existing literature that give the asymptotic behavior of conditional distributions similar to $\boldsymbol{Z}|\boldsymbol{N}(n)$. In response adaptive allocation designs the i.i.d. assumption is invalid, so there is minimal intuitive expectation that $\mbox{Var}[\boldsymbol{Z}|\boldsymbol{N}(n)]$ will converge to a constant matrix. This intuition is supported in the next section where it is show, heuristically, that $\mbox{Var}[\boldsymbol{Z}|\boldsymbol{N}(n)]$ does not converge to a constant matrix. This leaves us with the following question: if relative efficiency of conditional to unconditional information does not converge to a constant matrix, then is it appropriate to suggest that there is information lost by conditioning on the observed treatment sample sizes?


\subsubsection{Examples}

For the purpose of illustration consider one-half of the trace of the relative efficiency, Tr[Var$[\boldsymbol{Z}|\boldsymbol{N}(n)]$]/2. In optimal design the trace of the variance-covariance matrix is known as the $A$-optimality criterion. This criterion measures the average variance of the parameter estimates. The interpretation of Tr[Var$[\boldsymbol{Z}|\boldsymbol{N}(n)]$]/2 is that if it is greater than 1 then it indicates that the conditional information is greater than the unconditional information with respect to $A$-optimality. Figure \ref{fig:asyDist} presents smoothed histograms of the exact distribution of Tr[Var$[\boldsymbol{Z}|\boldsymbol{N}(n)]$]/2 for the RPW rule (top), the SDD (middle) and the NAD (bottom). In each row values of $p_{1} = p_{2} =$  0.5, 0.7 and 0.9 (left to right) were used. Within each figure each curve represents a different sample size, $n=25$ (dotted), $n=50$ (dashed) and $n=100$ (solid).

Consider the histograms for the RPW rule. The first notable feature is that the relative efficiency does not appear to converge to a constant, but instead converges to a non-degenerate distribution. The non-degeneracy of the relative efficiency has two implications. First, assessing the loss, or potentially gain, associated with conditional information is not straightforward, since it clearly depends on the observed treatment sample sizes. Second, the conditional observed and expected information measures do not appear to converge to a constant matrix. The figures also show that there exists non-zero probability that the conditional information is greater than the unconditional information. For $p_{1} = p_{2} = 0.5$ and 0.7 the majority of the probability is under 1.0. This is also true for values below 0.5 but is not shown here. However, for $p_{1} = p_{2} = 0.9$ there is significant probability that the relative efficiency will be greater than 1.0. These figures also provide insight into why \citet{Wei:Smyt:Lin:Stat:1990} concluded that unconditional inference is superior, since cases where both $p_{1}$ and $p_{2}$ were greater than 0.7 were not contained in the set of conditions considered in \citet{Wei:Smyt:Lin:Stat:1990}.

The figures for the SDD and the NAD also show a similar non-degenerate limiting behavior. For the SDD there is significant probability the relative efficiency is above 1.0 in all cases. This would indicate that conditional inference might be superior for a more substantial portion of the parameter space. This will be supported in the distribution study of Section \ref{sec:sim}. For the NAD, when $p_{1} = p_{2} = 0.5$ the distribution of the relative efficiency is highly right skewed and the mode is near 1.0. It might be expected that conditional inference will be better when the true value of the parameters are in the middle of the parameter space.


The main point of this section has been to demonstrate that conditioning does not \textit{always} represent information loss. For each of the 27 examples shown (3 samples for each of the 9 conditions) there is non-zero probability that conditional information is greater than unconditional information in 24 of them. Further, for each design there exist a subset of the parameter space where conditional information appears better with probability greater than one-half. For small and moderate sample sizes comparing the behavior of Var$[\boldsymbol{Z}|\boldsymbol{N}(n)]$ may not be and adequate description of the behavior of conditional and unconditional inference, which has greater practical implication than information. To better assess the small sample conditional inference properties the next section presents conditional estimation and inference.

\begin{figure}[!ht]
\centering
\begin{subfigure}{0.32\textwidth}
    \centering
    \includegraphics[scale=.41]{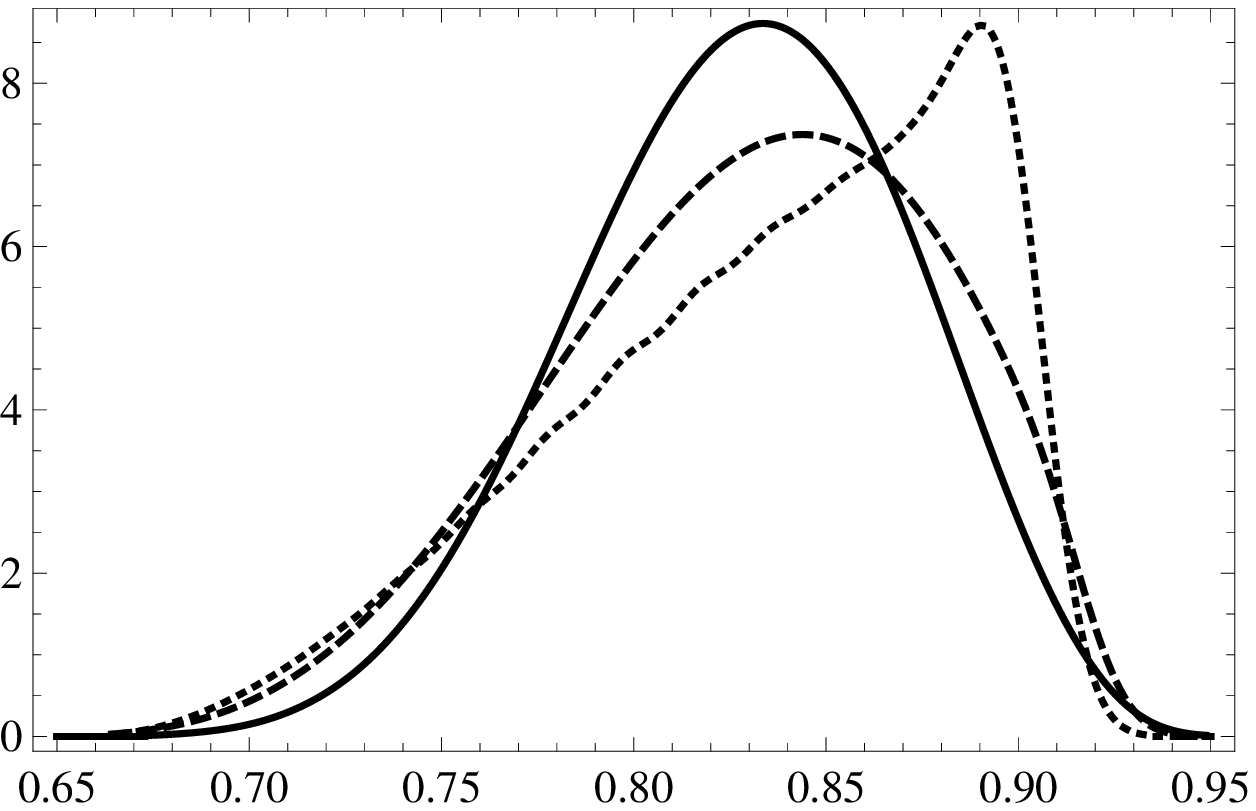}
\end{subfigure} %
\begin{subfigure}{0.32\textwidth}
    \centering
    \includegraphics[scale=.41]{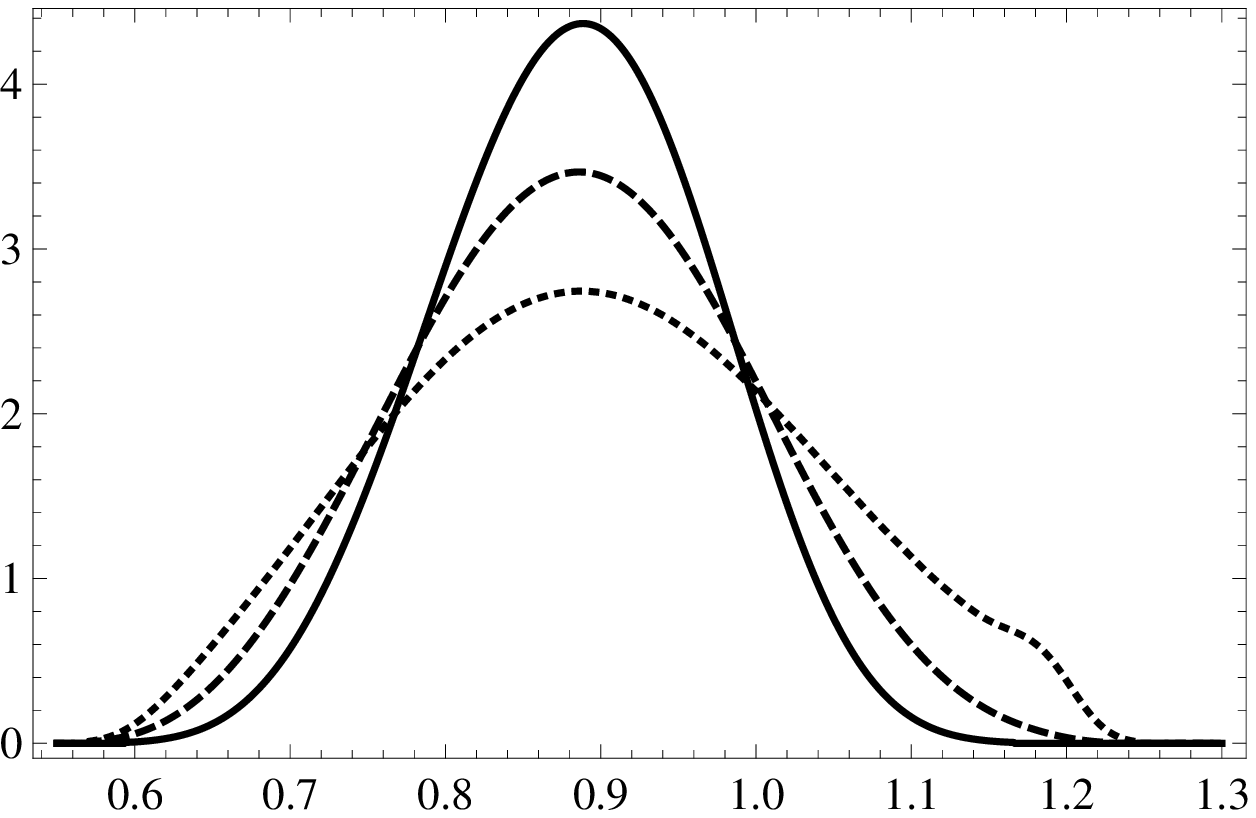}
\end{subfigure} %
\begin{subfigure}{0.32\textwidth}
    \centering
    \includegraphics[scale=.41]{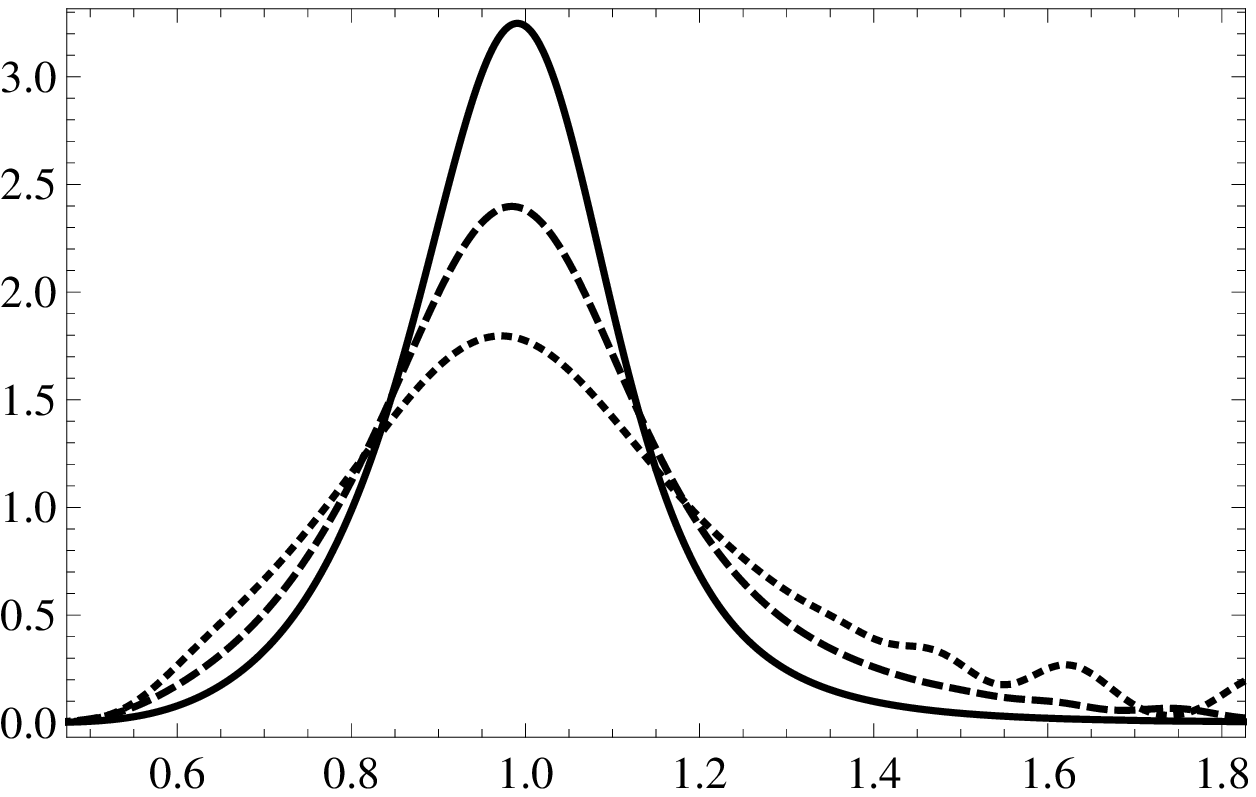}
\end{subfigure} \\
\begin{subfigure}{0.32\textwidth}
    \centering
    \includegraphics[scale=.41]{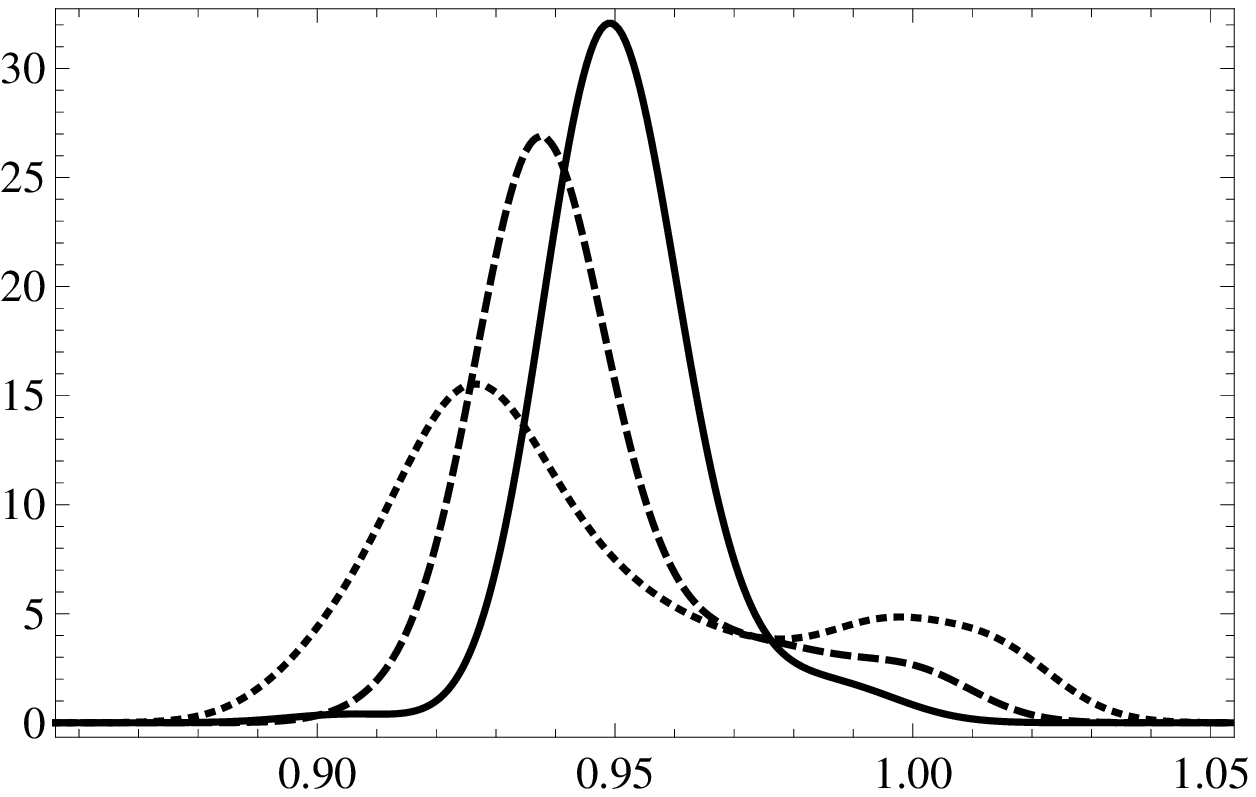}
\end{subfigure} %
\begin{subfigure}{0.32\textwidth}
    \centering
    \includegraphics[scale=.41]{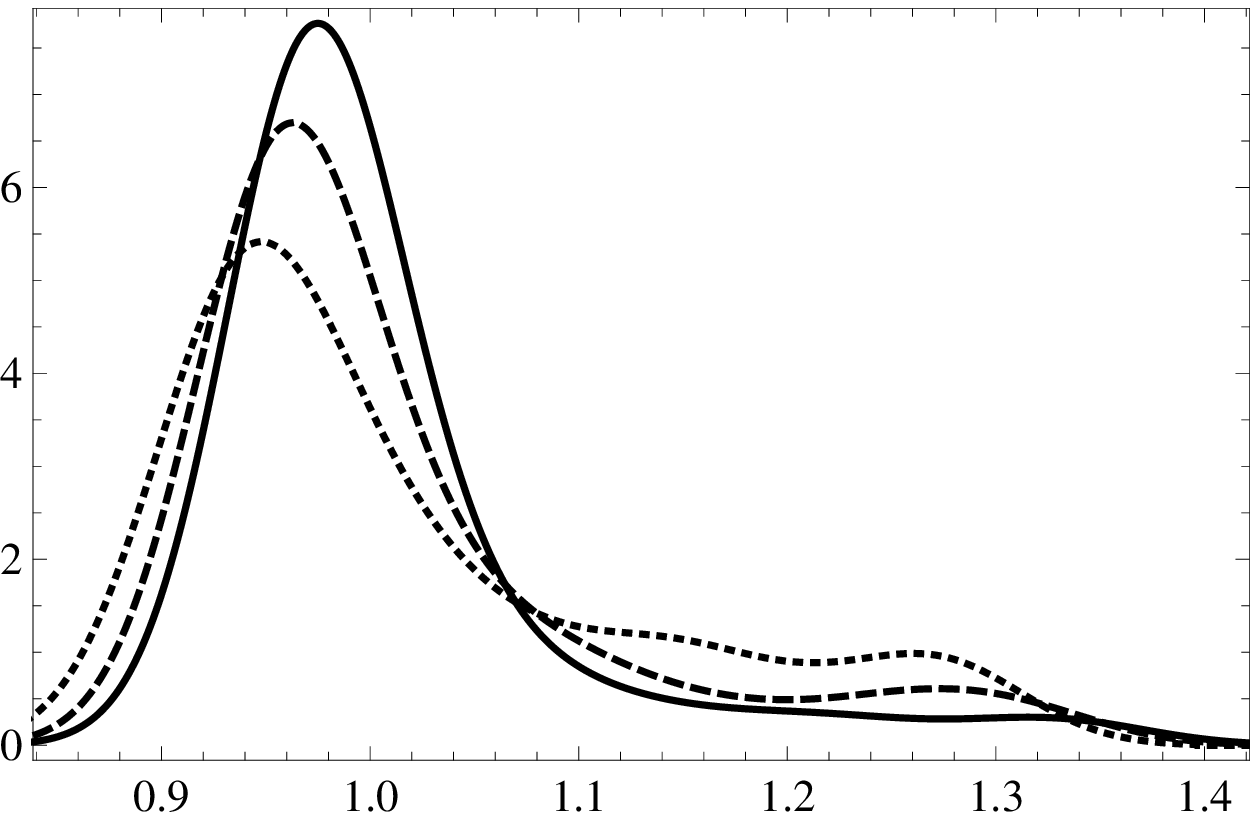}
\end{subfigure} %
\begin{subfigure}{0.32\textwidth}
    \centering
    \includegraphics[scale=.41]{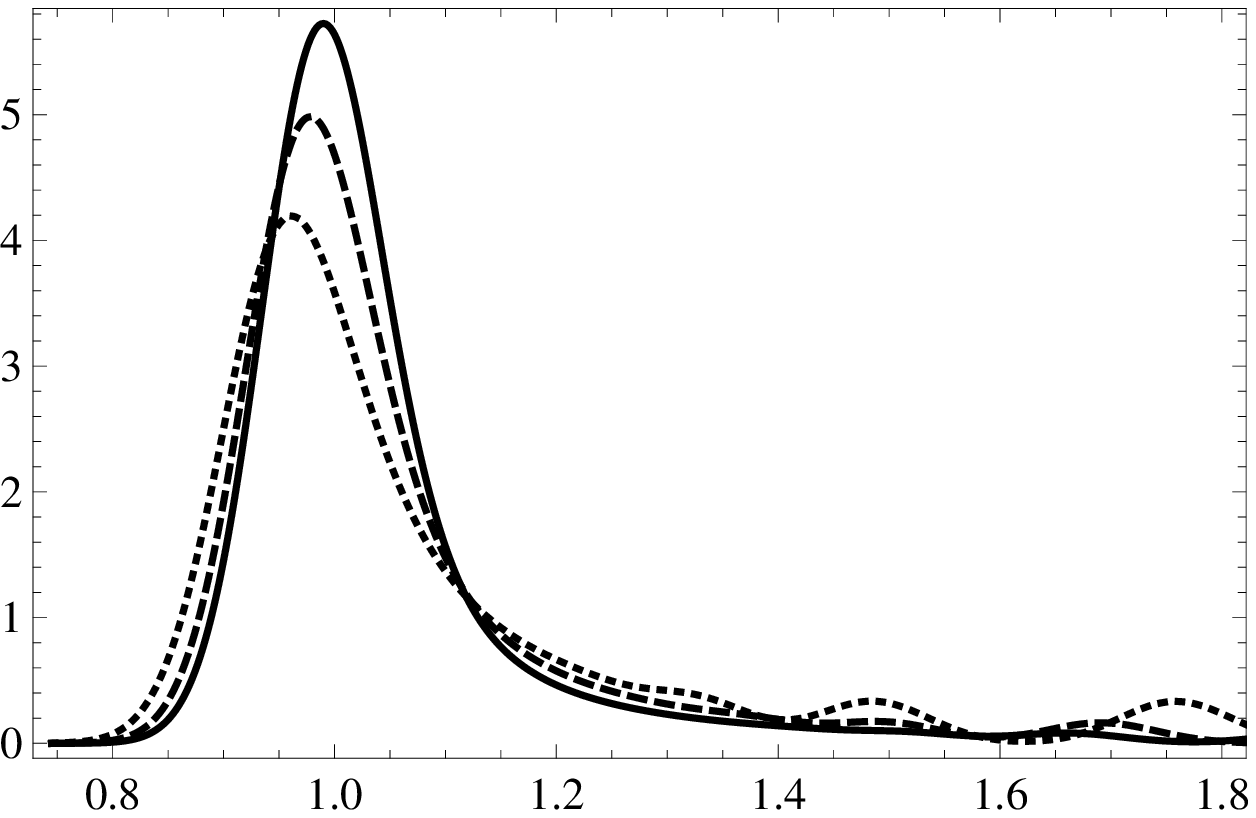}
\end{subfigure}\\
\begin{subfigure}{0.32\textwidth}
    \centering
    \includegraphics[scale=.41]{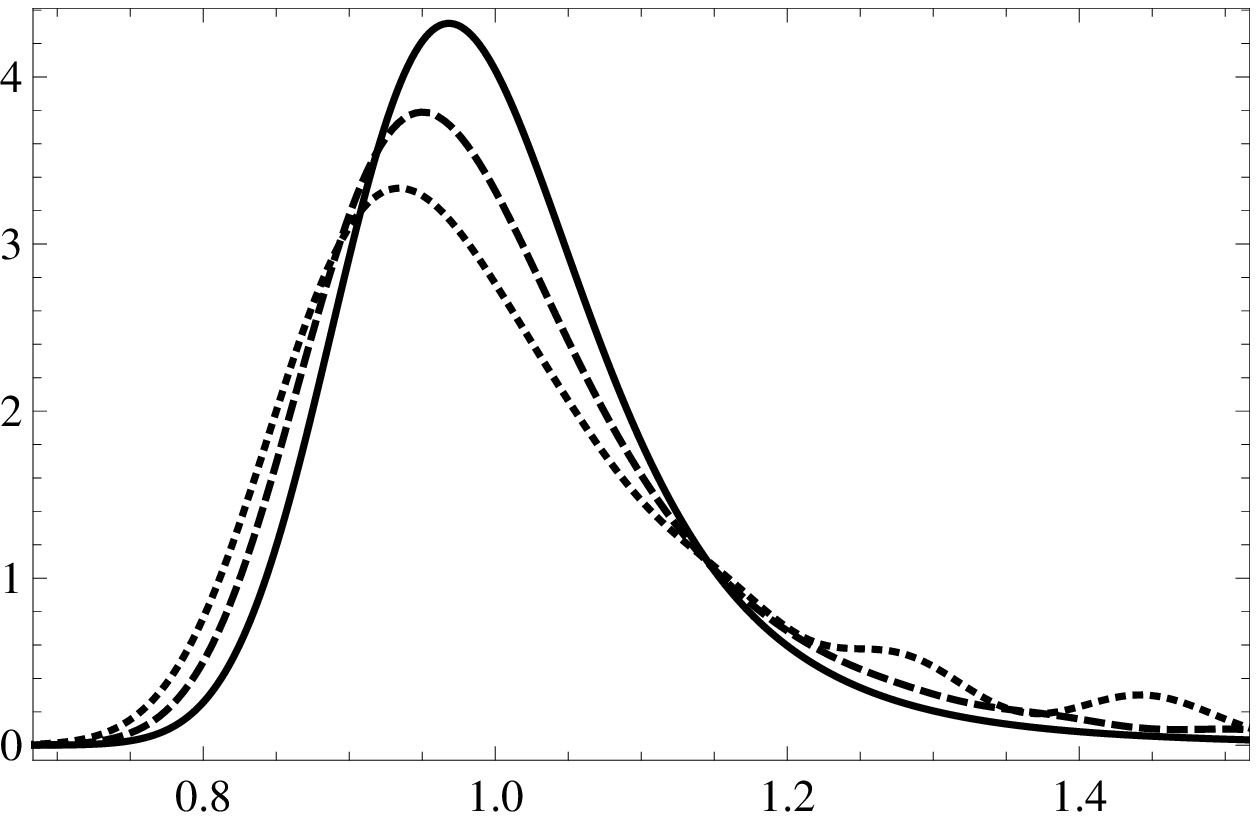}
\end{subfigure} %
\begin{subfigure}{0.32\textwidth}
    \centering
    \includegraphics[scale=.41]{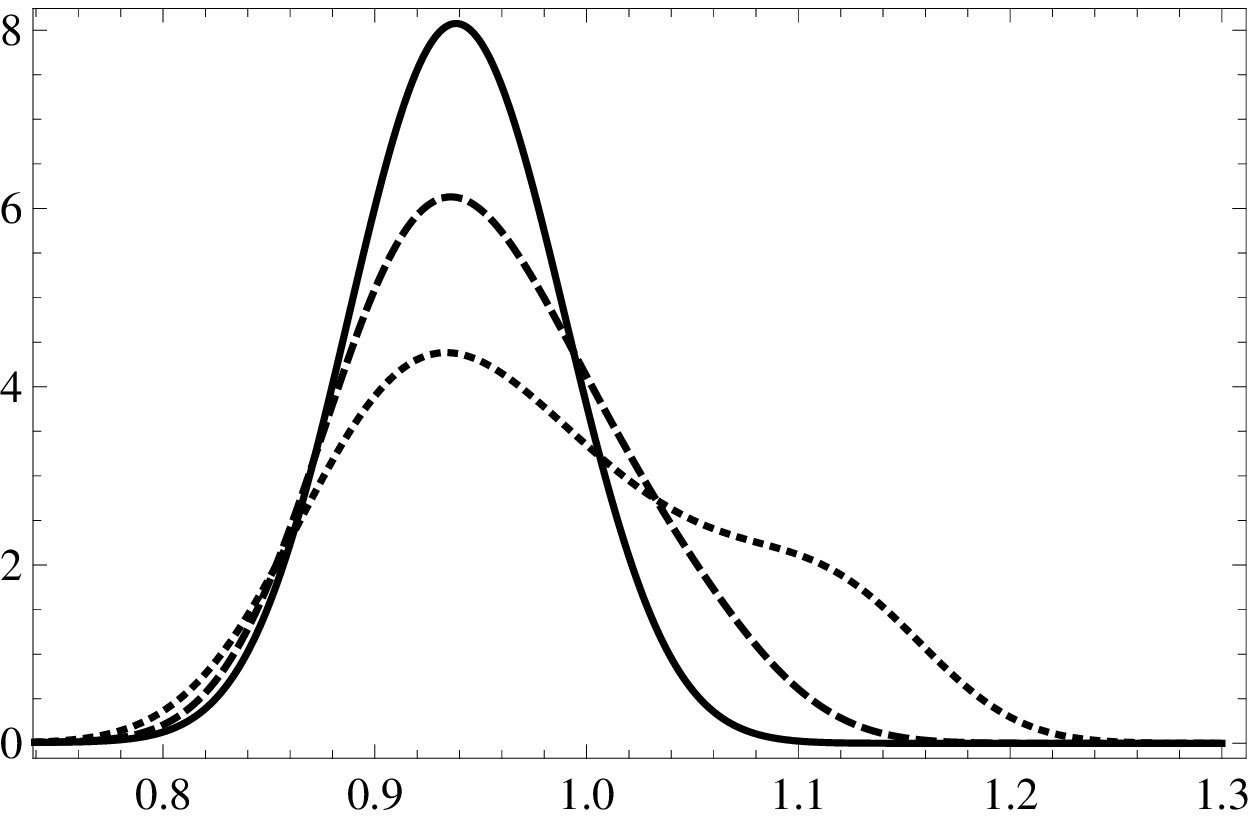}
\end{subfigure} %
\begin{subfigure}{0.32\textwidth}
    \centering
    \includegraphics[scale=.41]{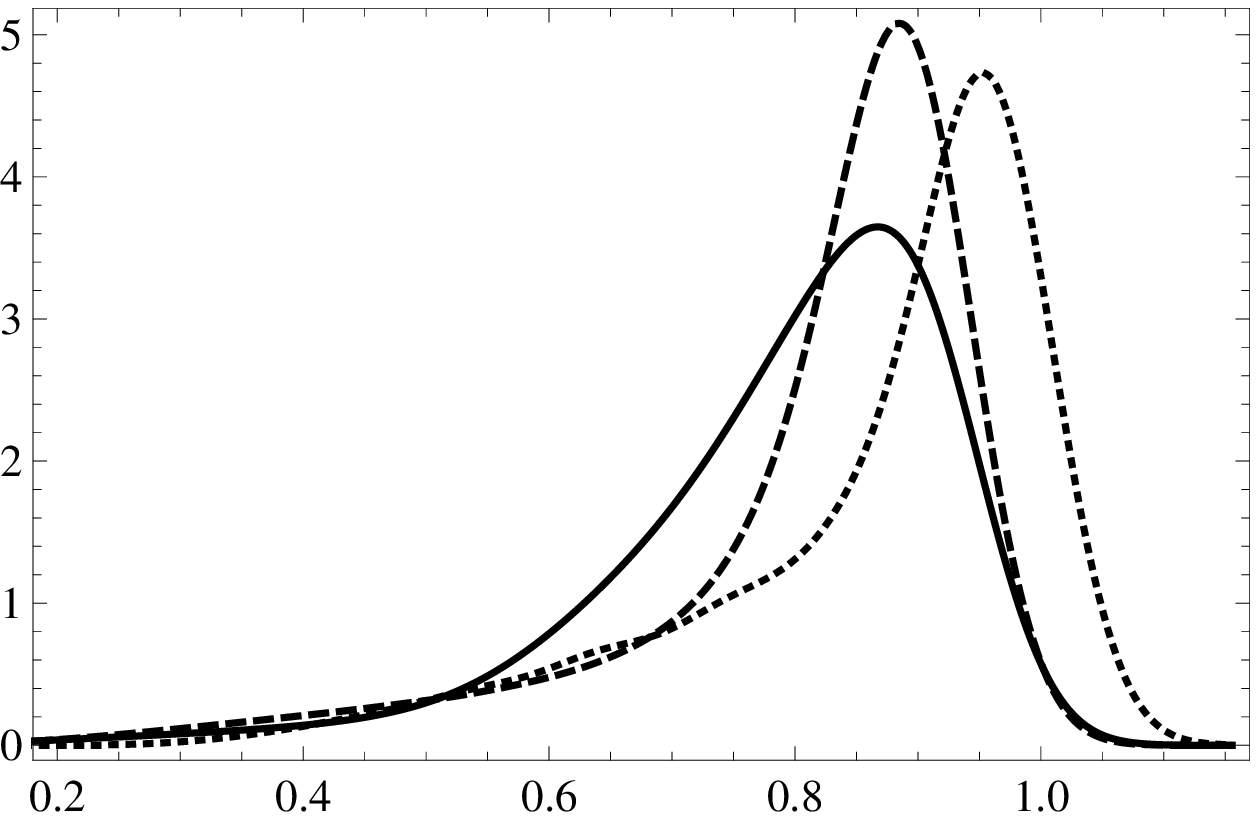}
\end{subfigure}
\caption{Smoothed histograms of the exact distribution of Var$[\boldsymbol{Z}|\boldsymbol{N}(n)]$ for the RPW (top), SDD (middle) and Neyman Allocation (bottom). In each row values of $p_{1} = p_{2} =$  0.5, 0.7 and 0.9 (left to right) were used. Within each figure each line represents a different sample size, $n=25$ (dotted), $n=50$ (dashed) and $n=100$ (solid). }\label{fig:asyDist}
\end{figure}

\section{Conditional Inference} \label{sec:Est}

The preceding section demonstrated that there can exist cases where the conditional information is greater than the unconditional information. If statistical inference is defined as assessing the precision, using confidence intervals or standard errors, of a relevant estimate, then this is a justification for the investigation of a conditional estimate and its corresponding confidence interval. In this section the conditional maximum likelihood estimate (CMLE) is considered.


\subsection{Estimation}

The CMLE is defined as
\begin{align} \label{eq:condMLE}
\boldsymbol{\tilde{p}} = \arg \min_{\boldsymbol{p}} L_{c}(\boldsymbol{p};\boldsymbol{s},\boldsymbol{n}).
\end{align}

\begin{theorem}\label{thm:condMLE}
Under condition \ref{C1} the solution to \eqref{eq:condMLE} is equivalent to the solution to
\begin{align} \label{eq:MOM}
\boldsymbol{\hat{p}} = {\rm{E}}\left[ \boldsymbol{\hat{p}}| \boldsymbol{N}(n) = \boldsymbol{n} \right] .
\end{align}
\end{theorem}
This theorem shows that the CMLE is also the conditional method of moments estimator. This form is convenient since because responses are dependent conventional maximum likelihood theory cannot be used directly to establish the asymptotic properties of the CMLE. Define the normalized CMLE as
\begin{align}
\boldsymbol{Z}_{\boldsymbol{H}} = \boldsymbol{H}^{1/2}(\boldsymbol{\tilde{p}} - \boldsymbol{p}),
\end{align}
where $\boldsymbol{H}$ can be either $I_{\boldsymbol{S}|\boldsymbol{N}}(\boldsymbol{\hat{p}})$ or $J_{\boldsymbol{S}|\boldsymbol{N}}(\boldsymbol{\hat{p}})$. For $\boldsymbol{Z}_{\boldsymbol{H}}$ the following can be shown.
\begin{theorem} \label{thm:condNorm}
Under conditions \ref{C1} and \ref{C2}
\begin{align}
\boldsymbol{\tilde{p}} - \boldsymbol{p} &\rightarrow 0  \\
\mbox{Var}[\boldsymbol{Z}_{\boldsymbol{H}}|\boldsymbol{N}(n)] &\rightarrow \boldsymbol{I}_{K}
\end{align}
in probability as $n\rightarrow\infty$.
\end{theorem}
This theorem establishes that the CMLE a is consistent estimate with a variance that is asymptotically equivalent to the inverses of $I_{\boldsymbol{S}|\boldsymbol{N}}(\boldsymbol{\hat{p}})$ and $J_{\boldsymbol{S}|\boldsymbol{N}}(\boldsymbol{\hat{p}})$.  The above theorem falls short of proving the asymptotic distribution of the normalized CMLE is standard multivariate normal. The difficulty in proving this limiting result was alluded to in the discussion of Figure \ref{fig:asyDist} where it was shown that it is not appropriate to assume that $n^{-1}I_{\boldsymbol{S}|\boldsymbol{N}}(\boldsymbol{\hat{p}})$ converges in probability to a constant matrix. However, an extensive simulation study was conducted (data not shown) and it was found that $\boldsymbol{Z}_{\boldsymbol{H}}$ does indeed appear to be asymptotically standard multivariate normally distributed.

A barrier to implementing the CMLE is the computational complexity associated with it. For example, using Theorem \ref{thm:condMLE} to find the MLE requires  $\mbox{E}\left[ \boldsymbol{\hat{p}}| \boldsymbol{N}(n) \right]$ to be computed exactly. This has no known solution, neither in general nor for the specific adaptive designs introduced in Section \ref{sec:Pre}. \citet{Wei:Exac:1988} developed a networking algorithm to generate the distribution of response-adaptive allocation designs. This procedure is not given here, instead see \citet{Wei:Exac:1988} or \citet{Ivan:Rose:Acom:2000} for details regarding this network algorithm. Once this distribution is found it is then required to solve for $\boldsymbol{p}$ using the equation \eqref{eq:condMLE} or \eqref{eq:MOM}.

\subsection{Confidence Intervals}

In this section a bootstrap confidence interval for the CMLE is developed. First, two unconditional confidence interval procedures are reviewed. Wald confidence intervals for $p_{k}$ of the form
\begin{align} \label{eq:Wald}
\hat{p}_{k}\pm z_{\alpha/2}\left\{\hat{\Lambda}_{\boldsymbol{n}}^{(kk)}\right\}^{-1/2},
\end{align}
where $\hat{\Lambda}_{\boldsymbol{n}}^{(kk)}$ is the $k$th diagonal element of $\hat{\Lambda}_{\boldsymbol{n}}$ and $z_{\alpha}$ is the $1-\alpha$ quantile of the standard normal distribution are asymptotically justified when the conditions of \citet{Rose:Flou:Durh:Asym:1997} or \citet{May:Flou:Asym:2009} are satisfied. However, \citet{Rose:Hu:Boot:1999} found \eqref{eq:Wald} had a tendency to be anti-conservative, as a result they developed a parametric bootstrap procedure, which they found nearly always outperformed \eqref{eq:Wald}. The \citet{Rose:Hu:Boot:1999} procedure will be referred to as the unconditional bootstrap procedure. Briefly, the unconditional bootstrap procedure is to, first, generate a bootstrap distribution by replicating the response adaptive allocation design using the observed success probabilities $\boldsymbol{\hat{p}}$ in place of the true success probabilities $\boldsymbol{p}$. For each bootstrap replication denote $\boldsymbol{S}^{*}$ and $\boldsymbol{N}^{*}$ as the vectors of successes and sample sizes observed in the replication.  Then the percentile bootstrap intervals are defined as the $\alpha/2$ and $(1-\alpha/2)$ quantiles of the bootstrap distribution of $\boldsymbol{\hat{p}}^{*}$.

The approximate normality of the CMLE has not been established and Wald confidence interval may not be appropriate. Instead, a conditional bootstrap procedure is proposed for confidence intervals for the CMLE. An obvious analog to the unconditional bootstrap procedure is to repeat the procedure finding $\boldsymbol{\tilde{p}}^{*}$ in place of $\boldsymbol{\hat{p}}^{*}$ in each bootstrap replication and then base the intervals on the relevant conditional quantiles of $\boldsymbol{\tilde{p}}^{*}$. However, the computation time required to compute $\boldsymbol{\tilde{p}}^{*}$ makes such a procedure impractical even for moderate sample sizes. To reduce computation time a conditional bootstrap procedure is developed that exploits the monotonicity of $\tilde{p}_{k}$ with respect to $S_{k}(n)$. The monotonicity relation is shown in Appendix \ref{app:Mono}. The importance of monotonicity is that, on the conditional space, the ordering of $S_{k}(n)$ corresponds to the ordering of $\tilde{p}_{k}$. The following modified conditional bootstrap procedure can be used to generate conditional confidence intervals for $p_{k}, k = 1,\ldots,K$.
\begin{enumerate}
\item{Generate $B$ replicates of the response adaptive allocation design using the observed success probabilities $\boldsymbol{\hat{p}}$ in place of the true success probabilities $\boldsymbol{p}$.}
\item{Compute $S_{k}^{*}$ and $\boldsymbol{N}^{*}$ for each replication. Delete bootstrap replicates such that $\boldsymbol{N}^{*}$ is not equal to $\boldsymbol{N}(n)$. This creates a conditional bootstrap distribution. Let $B_{C}$ denote the number of bootstrap samples satisfying the conditional requirement. }
\item{Let $S_{k}^{*(1)},\ldots,S_{k}^{*(B_{C})}$ be the ordered conditional bootstrap samples. Let $\boldsymbol{S}^{[k]}(n)$ be the vector of observed successes with $S_{k}(n)$ removed. }
\item{Let $\tilde{p}_{k}^{z}$ denote the CMLE calculated using $[\boldsymbol{S}^{[k]}(n),S_{k}^{*(B_{C}z)},\boldsymbol{N}(n)]$, where $z\in(0,1)$. The interval $(\tilde{p}_{k}^{\alpha/2},\tilde{p}_{k}^{1-\alpha/2})$ is the $1-\alpha$ two-sided conditional confidence interval for $p_{k}$.}
\end{enumerate}
The above bootstrap procedure creates intervals for each $p_{k}$, $k=1,\ldots,K$ separately. It is not necessary to repeat step 1 for each $p_{k}$ of interest. Finding the percentiles in terms of $S_{k}(n)$ and then calculating the CMLE at only the desired points avoids calculating $\tilde{p}_{k}^{*}$ for each bootstrap replicate. However, it should be understood that $\tilde{p}_{k}^{*}$ is not a monotonic function of $S_{j}(n)$, $j\ne k$. Therefore, the there may exist a trade-off in precision for reducing the computation time by not calculating $\tilde{p}_{k}^{*}$ for every replicate.

\subsection{Consistency of the Conditional Bootstrap}

As previously stated, heuristically it has been observed that the distribution of the normalized CMLE, $\boldsymbol{Z}_{\boldsymbol{H}}$, is approximately normally distributed. If this were true then the consistency of the conditional bootstrap would follow from approximately normality. However, it is desirable to avoid relying on this is unproven result to generate confidence intervals. The following theorem ensures the consistency of the conditional bootstrap distribution without relying on the approximately normality of $\boldsymbol{Z}_{\boldsymbol{H}}$.
\begin{theorem} \label{thm:boot}
Under conditions \ref{C1} and \ref{C2} the following holds
\begin{align}
P\{\sqrt{n}(\tilde{p}_{k} - p_{k}) \le \tilde{p}_{k}^{\alpha} | \boldsymbol{N}(n)=\boldsymbol{n} \} \longrightarrow \alpha.
\end{align}
in probability as $n\rightarrow\infty$.
\end{theorem}
Proof is in the appendix.

A remark on the width of the bootstrap confidence intervals. For the unconditional bootstrap procedure, if the asymptotic normality of the UMLE holds then the width of the unconditional  bootstrap intervals are asymptotically equivalent to $z_{\alpha/2}\mbox{Var}[\hat{p}_{i}]^{1/2}$. Similarly, if the CMLE is asymptotically normal then the average width of the conditional bootstrap intervals are asymptotically equivalent to $z_{\alpha/2}\mbox{E}[\mbox{Var}[\tilde{p}_{i}|\boldsymbol{N}(n)]^{1/2}]$. When comparing the relative efficiency of the variance of the two estimates it is more relevant to compare Var$[\hat{p}_{i}]$ to E$[Var[\tilde{p}_{i}|\boldsymbol{N}(n)]]$.

\section{Distribution Study} \label{sec:sim}

This section evaluates the performance of unconditional and conditional inference for the three examples given in Section \ref{sec:Pre}. Specifically, the following are reported: (1) the total absolute bias, TBias$(\boldsymbol{\hat{p}}) = \sum_{k} |E[\hat{p}_{k}] - p_{k}|$; (2) the trace of the variance of $\boldsymbol{\tilde{p}}$ relative to $\boldsymbol{\hat{p}}$
\begin{align}
\mbox{Rel-Var}(\boldsymbol{\tilde{p}},\boldsymbol{\hat{p}}) = \frac{\mbox{Tr}\{\mbox{Var}[\boldsymbol{\hat{p}}]\} }{\mbox{Tr}\{\mbox{E}[\mbox{Var}[\boldsymbol{\tilde{p}}|\boldsymbol{N}(n)]] \}};
\end{align}
(3) the relative length defined as
\begin{align}
\mbox{Rel-L}(\boldsymbol{\tilde{p}},\boldsymbol{\hat{p}}) = \frac{L_{\hat{p}_{1}} + L_{\hat{p}_{2}}}{L_{\tilde{p}_{1}} + L_{\tilde{p}_{2}}},
\end{align}
where $L_{\hat{p}_{k}}$ and $L_{\tilde{p}_{k}}$ are the the $95\%$ Bonferonni simultaneous unconditional and conditional intervals for $p_{i}, i=1,2$, respectively; and (4) the coverage of the of the simultaneous intervals, denoted C$_{\boldsymbol{\hat{p}}}$ and C$_{\boldsymbol{\tilde{p}}}$. When assessing the relative performance of the CMLE it is important to compare both the relative variance and the relative performance of the confidence intervals. The Rel-Var is a measure of the variance the CMLE relative to the variance of the UMLE. The Rel-L measures the efficiency of the proposed conditional bootstrap method relative to unconditional procedures. The interpretation for both Rel-Var and Rel-L is that values greater than 1 indicate that conditional inference is better for the respective measure.

Unconditional confidence intervals were computed using both the Wald and bootstrap procedures described in Section \ref{sec:Est}. Both are defined relative to the conditional bootstrap procedure proposed in that same section. The coverage of the Wald intervals and their length relative to the conditional bootstrap are denoted C$^{W}_{\boldsymbol{\hat{p}}}$ and Rel-L$^{W}(\boldsymbol{\tilde{p}},\boldsymbol{\hat{p}})$, respectively. The same quantities are denoted C$^{B}_{\boldsymbol{\hat{p}}}$ and Rel-L$^{B}(\boldsymbol{\tilde{p}},\boldsymbol{\hat{p}})$ for the unconditional bootstrap intervals.

For the RPW and the SDD $\alpha=\beta=1$ are used. For all 3 the designs results are given for $n=25, 50$, $p_{1}=0.1,0.3,0.5,0.7,0.9$, $p_{2}=0.1,0.3,0.5,0.7,0.9$ and $\alpha=0.05$. Note $p_{1}=a$ and $p_{2}=b$ gives the exact same results as $p_{1}=b$ and $p_{2}=a$ and thus duplicate entries are not included in the table. Cases where $\hat{p}_{k}=0,1$ or $N_{k}(n)=0$, $k=1,2$ were excluded. Tables \ref{tab:SDD}, \ref{tab:RPW} and \ref{tab:NAD} report the results for the SDD, RPW rule and NAD, respectively.

Figure \ref{fig:asyDist} indicated that conditional information was likely to be more beneficial in the SDD than for the RPW rule. For this reason, the results for the SDD, reported in Table \ref{tab:SDD}, are described first. For both sample sizes the Rel-Var$(\boldsymbol{\tilde{p}},\boldsymbol{\hat{p}})$ was greater than 1 for many of the parameter values considered. This is what was predicted in the discussion of Figure \ref{fig:asyDist} and indicates that conditional MLE is more efficient, with respect to the trace of variance, than the unconditional MLE for much of the parameter space; specifically, when both treatment success probabilities are greater than 0.5. Just as significant is that the Rel-L$^{B}(\boldsymbol{\tilde{p}},\boldsymbol{\hat{p}})$ is greater than 1 in every case except when $p_{1}=p_{2}=0.1$ and the Rel-L$^{W}(\boldsymbol{\tilde{p}},\boldsymbol{\hat{p}})$ was greater than 1 in every case. This implies that the SDD conditional confidence intervals are nearly uniformly narrower than the corresponding unconditional internals. For example when $n=50$ and $p_{1}=p_{2}=0.7$ the conditional intervals are on average nearly 1/2 the width of their unconditional bootstrap counterpart. Both the conditional and unconditional bootstrap confidence intervals provide coverage near the nominal level $(95\%)$ for most of the parameter space; however, they are conservative when the success probabilities are near 0 or 1. The unconditional Wald confidence intervals had a tendency to be unacceptably anti-conservative, as observed by \citet{Rose:Hu:Boot:1999}. The total absolute bias of $\boldsymbol{\tilde{p}}$ is uniformly less than the total absolute bias of $\boldsymbol{\hat{p}}$. 

Next, consider the RPW rule reported in Table \ref{tab:RPW}. There are many cases where Rel-Var$(\boldsymbol{\tilde{p}},\boldsymbol{\hat{p}})$, Rel-L$^{B}(\boldsymbol{\tilde{p}},\boldsymbol{\hat{p}})$ and Rel-L$^{W}(\boldsymbol{\tilde{p}},\boldsymbol{\hat{p}})$ are greater than 1; however, fewer cases were found than for the SDD. Once again conditional inference performed better as the success probabilities increase, in particular when both $p_{1}$ and $p_{2}$ are greater than 0.7. This agrees with what was observed in Figure \ref{fig:asyDist}. It is of interest to note that \citet{Smyt:Rose:Play:1995} show that there is a phase transition such that $N_{k}(n)$, normalized by its mean and variance, is asymptotically normally distributed only when $p_{1}+p_{2}<1.5$. There may be a connection with this phase transition and the conditional information; however, this has not been investigated theoretically. Once again the coverage of both bootstrap intervals was adequate; but the coverage of the Wald intervals were significantly anti-conservative in many cases. For the RPW rule the total absolute bias of $\boldsymbol{\tilde{p}}$ is uniformly less than that of $\boldsymbol{\hat{p}}$.

Finally, consider the NAD, reported in Table \ref{tab:NAD}. The Rel-Var$(\boldsymbol{\tilde{p}},\boldsymbol{\hat{p}})$ was greater than 1 when the success probabilities were both between 0.3 and 0.7. When $n=25$ the Rel-L$^{B}(\boldsymbol{\tilde{p}},\boldsymbol{\hat{p}})$ and the Rel-L$^{W}(\boldsymbol{\tilde{p}},\boldsymbol{\hat{p}})$ were greater than 1 in all but a single case. However, when $n$ was increased to 50 the relative length of the conditional and unconditional bootstrap confidence intervals was fairly balanced.

In summary. For each design and each sample size there are cases where the variance of the CMLE was less than the variance of the UMLE, at least with respect to $A$-optimality. Further, in the 90 different cases presented (3 designs, 2 samples sizes and 15 parameter conditions) the length of the conditional bootstrap confidence intervals was at least as narrow as the corresponding unconditional bootstrap confidence intervals in 66 cases. The conditional bootstrap confidence intervals was at least as narrow as the corresponding unconditional Wald confidence intervals in 80 out of 90 cases. Overall both of the bootstrap confidence interval procedures exhibited adequate coverage. The Wald confidence intervals exhibited a concerning tendency to be anti-conservative. The bias of the CMLE was uniformly better than the UMLE. This simulation study demonstrates that there is significant justification for the use of conditional inference in response adaptive allocation designs.

The same distribution study was conducted for designs with the objective of optimizing the variance with respect to the simple difference, relative risk and odds ratio [see \citet{Rose:Lach:Rand:2002} ch. 10]. For each of these designs the findings were similar to designs presented in this section. There were always cases where conditional inference was superior to unconditional and visa versa. Tables for these designs are presented in the supplementary materials. A significance level of $\alpha=0.10$ was also considered; however, the the results were similar to what is described and are not shown.

One pattern that is present in all cases was that the conditional confidence intervals had greater relative efficiency than the relative efficiency of the conditional variance. Specifically, Rel-L$^{B}(\boldsymbol{\tilde{p}},\boldsymbol{\hat{p}})$ was always greater than Rel-Var$(\boldsymbol{\tilde{p}},\boldsymbol{\hat{p}})$. The true cause of this observation is unknown; however, it is speculated it is, at least in part, caused by the ordering problem discussed in \citet{Begg:OnIn:1990}. The \citet{Rose:Hu:Boot:1999} procedure orders the bootstrap distribution only with respect to $\hat{p}_{k}^{*}$.  Since, on the unconditional space $\hat{p}_{k}$ is not a sufficient statistic the simple ordering interval does not fully incorporate the observed sample sizes. If an ordering scheme could be developed that incorporated both $\hat{p}_{k}$ and the observed sample size it is possible that this could increase the efficiency of the unconditional bootstrap procedure for small and moderate sample sizes.

\begin{table}[!ht]
\centering
\small
\begin{tabular}{ cccccccccc }
$p_{1}$ & $p_{2}$ & TBias$(\boldsymbol{\tilde{p}})$ & TBias$(\boldsymbol{\hat{p}})$ & Rel-Var$(\boldsymbol{\tilde{p}},\boldsymbol{\hat{p}})$ &  Rel-L$^{B}(\boldsymbol{\tilde{p}},\boldsymbol{\hat{p}})$ & Rel-L$^{W}(\boldsymbol{\tilde{p}},\boldsymbol{\hat{p}})$ & C$_{\boldsymbol{\tilde{p}}}$ & C$^{B}_{\boldsymbol{\hat{p}}}$ & C$^{W}_{\boldsymbol{\hat{p}}}$ \\
\noalign{\smallskip}\hline\hline\noalign{\smallskip}
\multicolumn{10}{c}{$n=25$} \\
\noalign{\smallskip}\hline\noalign{\smallskip}
0.1 & 0.1 & 0.01 & 0.03 & 0.69 & 0.93 & 1.11 & 0.9977 & 0.9987 & 0.9960 \\
0.3 & 0.1 & 0.02 & 0.04 & 0.80 & 1.06 & 1.08 & 0.9709 & 0.9706 & 0.9479 \\
0.3 & 0.3 & 0.03 & 0.06 & 0.88 & 1.20 & 1.06 & 0.9597 & 0.9868 & 0.9315 \\
0.5 & 0.1 & 0.02 & 0.03 & 0.84 & 1.35 & 1.07 & 0.9555 & 0.9603 & 0.9370 \\
0.5 & 0.3 & 0.04 & 0.07 & 0.94 & 1.42 & 1.06 & 0.9550 & 0.9811 & 0.9228 \\
0.5 & 0.5 & 0.04 & 0.08 & 1.04 & 1.50 & 1.06 & 0.9410 & 0.9933 & 0.9089 \\
0.7 & 0.1 & 0.02 & 0.03 & 0.84 & 1.53 & 1.10 & 0.9576 & 0.9636 & 0.9410 \\
0.7 & 0.3 & 0.04 & 0.06 & 0.95 & 1.50 & 1.08 & 0.9495 & 0.9738 & 0.9255 \\
0.7 & 0.5 & 0.04 & 0.07 & 1.08 & 1.53 & 1.07 & 0.9350 & 0.9912 & 0.9095 \\
0.7 & 0.7 & 0.04 & 0.07 & 1.17 & 1.59 & 1.07 & 0.9399 & 0.9977 & 0.9069 \\
0.9 & 0.1 & 0.02 & 0.03 & 0.79 & 1.47 & 1.17 & 0.9876 & 0.9976 & 0.9918 \\
0.9 & 0.3 & 0.04 & 0.06 & 0.91 & 1.39 & 1.14 & 0.9805 & 0.9978 & 0.9780 \\
0.9 & 0.5 & 0.04 & 0.06 & 1.05 & 1.42 & 1.11 & 0.9663 & 0.9994 & 0.9605 \\
0.9 & 0.7 & 0.03 & 0.05 & 1.17 & 1.55 & 1.10 & 0.9643 & 0.9998 & 0.9547 \\
0.9 & 0.9 & 0.02 & 0.03 & 1.22 & 1.69 & 1.10 & 0.9699 & 1.0000 & 0.9905 \\
\noalign{\smallskip}\hline\noalign{\smallskip}
\multicolumn{10}{c}{$n=50$}\\
\noalign{\smallskip}\hline\noalign{\smallskip}
0.1 & 0.1 & 0.01 & 0.02 & 0.69 & 0.89 & 1.02 & 0.9961 & 0.9994 & 0.9900 \\
0.3 & 0.1 & 0.01 & 0.03 & 0.75 & 1.01 & 1.02 & 0.9632 & 0.9360 & 0.9414 \\
0.3 & 0.3 & 0.02 & 0.05 & 0.89 & 1.14 & 1.01 & 0.9521 & 0.9263 & 0.9124 \\
0.5 & 0.1 & 0.01 & 0.03 & 0.73 & 1.07 & 1.05 & 0.9634 & 0.9610 & 0.9561 \\
0.5 & 0.3 & 0.02 & 0.06 & 0.92 & 1.26 & 1.03 & 0.9488 & 0.9540 & 0.9228 \\
0.5 & 0.5 & 0.03 & 0.07 & 1.07 & 1.52 & 1.02 & 0.9383 & 0.9599 & 0.9124 \\
0.7 & 0.1 & 0.01 & 0.03 & 0.71 & 1.22 & 1.09 & 0.9638 & 0.9685 & 0.9584 \\
0.7 & 0.3 & 0.03 & 0.06 & 0.88 & 1.40 & 1.05 & 0.9524 & 0.9646 & 0.9330 \\
0.7 & 0.5 & 0.03 & 0.06 & 1.10 & 1.70 & 1.04 & 0.9378 & 0.9686 & 0.9150 \\
0.7 & 0.7 & 0.02 & 0.06 & 1.24 & 1.94 & 1.04 & 0.9276 & 0.9762 & 0.9118 \\
0.9 & 0.1 & 0.02 & 0.03 & 0.69 & 1.65 & 1.13 & 0.9515 & 0.9556 & 0.9400 \\
0.9 & 0.3 & 0.03 & 0.06 & 0.84 & 1.52 & 1.09 & 0.9544 & 0.9759 & 0.9270 \\
0.9 & 0.5 & 0.03 & 0.06 & 1.04 & 1.55 & 1.07 & 0.9458 & 0.9883 & 0.9156 \\
0.9 & 0.7 & 0.02 & 0.05 & 1.23 & 1.77 & 1.06 & 0.9451 & 0.9966 & 0.9180 \\
0.9 & 0.9 & 0.01 & 0.02 & 1.31 & 1.89 & 1.10 & 0.9668 & 0.9997 & 0.9430 \\
\noalign{\smallskip}\hline\noalign{\smallskip}
\end{tabular}
\caption{Exact results for the total absolute bias, relative variance, relative length and coverage of the confidence intervals, for $\boldsymbol{\tilde{p}}$ and $\boldsymbol{\hat{p}}$ for the SDD. Values $n=25, 50$, $p_{1}=0.1,0.3,0.5,0.7,0.9$, $p_{2}=0.1,0.3,0.5,0.7,0.9$ and $\alpha=0.05$ were used.   } \label{tab:SDD}
\end{table}
\begin{table}[!ht]
\centering
\small
\begin{tabular}{ cccccccccc }
\noalign{\smallskip}\hline\hline\noalign{\smallskip}
$p_{1}$ & $p_{2}$ & TBias$(\boldsymbol{\tilde{p}})$ & TBias$(\boldsymbol{\hat{p}})$ & Rel-Var$(\boldsymbol{\tilde{p}},\boldsymbol{\hat{p}})$ &  Rel-L$^{B}(\boldsymbol{\tilde{p}},\boldsymbol{\hat{p}})$ & Rel-L$^{W}(\boldsymbol{\tilde{p}},\boldsymbol{\hat{p}})$ & C$_{\boldsymbol{\tilde{p}}}$ & C$^{B}_{\boldsymbol{\hat{p}}}$ & C$^{W}_{\boldsymbol{\hat{p}}}$ \\
\noalign{\smallskip}\hline\noalign{\smallskip}
\multicolumn{10}{c}{$n=25$}\\
\noalign{\smallskip}\hline\noalign{\smallskip}
0.1 & 0.1 & 0.01 & 0.01 & 0.77 & 0.90 & 1.05 & 0.9963 & 0.9984 & 0.9961 \\
0.3 & 0.1 & 0.01 & 0.01 & 0.80 & 0.92 & 1.02 & 0.9697 & 0.9608 & 0.9322 \\
0.3 & 0.3 & 0.00 & 0.02 & 0.84 & 0.96 & 1.00 & 0.9561 & 0.9693 & 0.8973 \\
0.5 & 0.1 & 0.00 & 0.02 & 0.81 & 0.93 & 1.01 & 0.9701 & 0.9654 & 0.9407 \\
0.5 & 0.3 & 0.00 & 0.03 & 0.87 & 0.99 & 1.00 & 0.9632 & 0.9627 & 0.9084 \\
0.5 & 0.5 & 0.01 & 0.03 & 0.92 & 1.04 & 1.00 & 0.9505 & 0.9407 & 0.8868 \\
0.7 & 0.1 & 0.00 & 0.02 & 0.79 & 0.96 & 1.03 & 0.9769 & 0.9637 & 0.9492 \\
0.7 & 0.3 & 0.01 & 0.03 & 0.86 & 1.03 & 1.02 & 0.9670 & 0.9615 & 0.9271 \\
0.7 & 0.5 & 0.01 & 0.04 & 0.95 & 1.11 & 1.01 & 0.9470 & 0.9503 & 0.8964 \\
0.7 & 0.7 & 0.02 & 0.04 & 1.04 & 1.22 & 1.02 & 0.9534 & 0.9459 & 0.8946 \\
0.9 & 0.1 & 0.01 & 0.02 & 0.75 & 1.16 & 1.10 & 0.9870 & 0.9880 & 0.9911 \\
0.9 & 0.3 & 0.03 & 0.05 & 0.84 & 1.17 & 1.09 & 0.9827 & 0.9890 & 0.9809 \\
0.9 & 0.5 & 0.03 & 0.05 & 0.96 & 1.21 & 1.07 & 0.9697 & 0.9888 & 0.9614 \\
0.9 & 0.7 & 0.03 & 0.05 & 1.10 & 1.31 & 1.06 & 0.9731 & 0.9831 & 0.9520 \\
0.9 & 0.9 & 0.02 & 0.03 & 1.25 & 1.40 & 1.07 & 0.9685 & 0.9955 & 0.9893 \\
\noalign{\smallskip}\hline\noalign{\smallskip}
\multicolumn{8}{c}{$n=50$}\\
\noalign{\smallskip}\hline\noalign{\smallskip}
0.1 & 0.1 & 0.00 & 0.00 & 0.83 & 0.91 & 0.99 & 0.9939 & 0.9962 & 0.9964 \\
0.3 & 0.1 & 0.00 & 0.01 & 0.82 & 0.92 & 0.96 & 0.9726 & 0.9597 & 0.9529 \\
0.3 & 0.3 & 0.00 & 0.01 & 0.86 & 0.94 & 0.95 & 0.9490 & 0.9232 & 0.9068 \\
0.5 & 0.1 & 0.00 & 0.01 & 0.80 & 0.91 & 0.96 & 0.9810 & 0.9710 & 0.9613 \\
0.5 & 0.3 & 0.00 & 0.01 & 0.87 & 0.95 & 0.95 & 0.9512 & 0.9326 & 0.9094 \\
0.5 & 0.5 & 0.00 & 0.02 & 0.90 & 0.98 & 0.95 & 0.9564 & 0.9426 & 0.9165 \\
0.7 & 0.1 & 0.00 & 0.01 & 0.79 & 0.92 & 0.98 & 0.9839 & 0.9693 & 0.9633 \\
0.7 & 0.3 & 0.00 & 0.02 & 0.86 & 0.97 & 0.97 & 0.9529 & 0.9371 & 0.9108 \\
0.7 & 0.5 & 0.00 & 0.02 & 0.92 & 1.02 & 0.96 & 0.9510 & 0.9387 & 0.9130 \\
0.7 & 0.7 & 0.01 & 0.02 & 0.98 & 1.08 & 0.97 & 0.9459 & 0.9342 & 0.9150 \\
0.9 & 0.1 & 0.01 & 0.01 & 0.71 & 0.98 & 1.06 & 0.9709 & 0.9533 & 0.9553 \\
0.9 & 0.3 & 0.01 & 0.03 & 0.81 & 1.05 & 1.04 & 0.9610 & 0.9563 & 0.9236 \\
0.9 & 0.5 & 0.01 & 0.04 & 0.93 & 1.18 & 1.02 & 0.9584 & 0.9600 & 0.9019 \\
0.9 & 0.7 & 0.01 & 0.03 & 1.09 & 1.31 & 1.01 & 0.9567 & 0.9575 & 0.9099 \\
0.9 & 0.9 & 0.01 & 0.02 & 1.29 & 1.36 & 1.04 & 0.9741 & 0.9794 & 0.9416 \\
\noalign{\smallskip}\hline\noalign{\smallskip}
\end{tabular}
\caption{Exact results for the total absolute bias, relative variance, relative length and coverage of the confidence intervals, for $\boldsymbol{\tilde{p}}$ and $\boldsymbol{\hat{p}}$ for the RPW. Values $n=25, 50$, $p_{1}=0.1,0.3,0.5,0.7,0.9$, $p_{2}=0.1,0.3,0.5,0.7,0.9$ and $\alpha=0.05$ were used.  } \label{tab:RPW}
\end{table}
\begin{table}[!ht]
\centering
\small
\begin{tabular}{ cccccccccc }
\noalign{\smallskip}\hline\hline\noalign{\smallskip}
$p_{1}$ & $p_{2}$ & TBias$(\boldsymbol{\tilde{p}})$ & TBias$(\boldsymbol{\hat{p}})$ & Rel-Var$(\boldsymbol{\tilde{p}},\boldsymbol{\hat{p}})$ &  Rel-L$^{B}(\boldsymbol{\tilde{p}},\boldsymbol{\hat{p}})$ & Rel-L$^{W}(\boldsymbol{\tilde{p}},\boldsymbol{\hat{p}})$ & C$_{\boldsymbol{\tilde{p}}}$ & C$^{B}_{\boldsymbol{\hat{p}}}$ & C$^{W}_{\boldsymbol{\hat{p}}}$ \\
\noalign{\smallskip}\hline\noalign{\smallskip}
\multicolumn{10}{c}{$n=25$}\\
\noalign{\smallskip}\hline\noalign{\smallskip}
0.1 & 0.1 & 0.01 & 0.01 & 0.89 & 1.01 & 1.15 & 0.9951 & 0.9966 & 0.9954 \\
0.3 & 0.1 & 0.01 & 0.01 & 0.97 & 1.02 & 1.08 & 0.9520 & 0.9607 & 0.9378 \\
0.3 & 0.3 & 0.01 & 0.01 & 1.03 & 1.05 & 1.04 & 0.9268 & 0.9539 & 0.8967 \\
0.5 & 0.1 & 0.00 & 0.01 & 0.99 & 1.03 & 1.06 & 0.9583 & 0.9557 & 0.9301 \\
0.5 & 0.3 & 0.00 & 0.01 & 1.04 & 1.06 & 1.02 & 0.9275 & 0.9425 & 0.8860 \\
0.5 & 0.5 & 0.00 & 0.00 & 1.06 & 1.07 & 1.00 & 0.9257 & 0.9255 & 0.8710 \\
0.7 & 0.1 & 0.01 & 0.01 & 0.97 & 1.02 & 1.08 & 0.9636 & 0.9607 & 0.9378 \\
0.7 & 0.3 & 0.01 & 0.01 & 1.03 & 1.06 & 1.04 & 0.9269 & 0.9539 & 0.8967 \\
0.7 & 0.5 & 0.00 & 0.01 & 1.04 & 1.06 & 1.02 & 0.9265 & 0.9425 & 0.8860 \\
0.7 & 0.7 & 0.01 & 0.01 & 1.03 & 1.06 & 1.04 & 0.9248 & 0.9539 & 0.8967 \\
0.9 & 0.1 & 0.01 & 0.01 & 0.89 & 0.98 & 1.12 & 0.9886 & 0.9966 & 0.9954 \\
0.9 & 0.3 & 0.01 & 0.01 & 0.97 & 1.03 & 1.08 & 0.9532 & 0.9607 & 0.9378 \\
0.9 & 0.5 & 0.00 & 0.01 & 0.99 & 1.04 & 1.07 & 0.9571 & 0.9557 & 0.9301 \\
0.9 & 0.7 & 0.01 & 0.01 & 0.97 & 1.03 & 1.09 & 0.9534 & 0.9607 & 0.9378 \\
0.9 & 0.9 & 0.01 & 0.01 & 0.89 & 1.00 & 1.14 & 0.9899 & 0.9966 & 0.9954 \\
\noalign{\smallskip}\hline\noalign{\smallskip}
\multicolumn{10}{c}{$n=50$}\\
\noalign{\smallskip}\hline\noalign{\smallskip}
0.1 & 0.1 & 0.00 & 0.01 & 0.85 & 0.96 & 1.04 & 0.9954 & 0.9969 & 0.9954 \\
0.3 & 0.1 & 0.00 & 0.01 & 0.92 & 0.99 & 1.02 & 0.9597 & 0.9540 & 0.9500 \\
0.3 & 0.3 & 0.00 & 0.01 & 1.01 & 1.04 & 1.00 & 0.9268 & 0.9122 & 0.9021 \\
0.5 & 0.1 & 0.00 & 0.01 & 0.93 & 0.99 & 1.02 & 0.9639 & 0.9671 & 0.9567 \\
0.5 & 0.3 & 0.00 & 0.00 & 1.02 & 1.04 & 1.00 & 0.9310 & 0.9222 & 0.9069 \\
0.5 & 0.5 & 0.00 & 0.00 & 1.04 & 1.05 & 1.00 & 0.9349 & 0.9359 & 0.9116 \\
0.7 & 0.1 & 0.00 & 0.01 & 0.92 & 1.00 & 1.03 & 0.9601 & 0.9540 & 0.9500 \\
0.7 & 0.3 & 0.00 & 0.01 & 1.01 & 1.04 & 1.00 & 0.9267 & 0.9122 & 0.9021 \\
0.7 & 0.5 & 0.00 & 0.00 & 1.02 & 1.04 & 1.00 & 0.9307 & 0.9222 & 0.9069 \\
0.7 & 0.7 & 0.00 & 0.01 & 1.01 & 1.04 & 1.00 & 0.9267 & 0.9122 & 0.9021 \\
0.9 & 0.1 & 0.00 & 0.01 & 0.85 & 0.97 & 1.05 & 0.9938 & 0.9969 & 0.9954 \\
0.9 & 0.3 & 0.00 & 0.01 & 0.92 & 1.00 & 1.03 & 0.9588 & 0.9540 & 0.9500 \\
0.9 & 0.5 & 0.00 & 0.01 & 0.93 & 0.99 & 1.02 & 0.9639 & 0.9671 & 0.9567 \\
0.9 & 0.7 & 0.00 & 0.01 & 0.92 & 1.00 & 1.03 & 0.9589 & 0.9540 & 0.9500 \\
0.9 & 0.9 & 0.00 & 0.01 & 0.85 & 0.97 & 1.05 & 0.9944 & 0.9969 & 0.9954 \\
\noalign{\smallskip}\hline\noalign{\smallskip}
\end{tabular}
\caption{Exact results for the total absolute bias, relative variance, relative length and coverage of the confidence intervals, for $\boldsymbol{\tilde{p}}$ and $\boldsymbol{\hat{p}}$ for the NAD. Values $n=25, 50$, $p_{1}=0.1,0.3,0.5,0.7,0.9$, $p_{2}=0.1,0.3,0.5,0.7,0.9$ and $\alpha=0.05$ were used. } \label{tab:NAD}
\end{table}

\section{Real Data Example} \label{sec:real}

We illustrate the performance of conditional and unconditional inference for the fluoxetine-placebo controlled clinical trial. The study along with extensive details can be found in \citet{Tamu:Fari:Ande:Heil:ACas:1994}. \citet{Rose:Hu:Boot:1999} also consider this example. In the trial patients were stratified into two groups, those with normal rapid eye movement latency (REML) and those with shortened REML. Randomization took place separately within strata. The trial initialized with the first six patients in each strata assigned to treatment according to a permuted block randomization. After this initialization a RPW(1,1) was used for the duration, once again separately for each strata. The RPW allocation rule was updated based on defining a success as a greater than 50$\%$ reduction in the Hamilton Depression Scale in two consecutive visits at least three weeks post therapy. Note this was considered a surrogate marker for the primary endpoint 50$\%$ reduction in the Hamilton Depression Scale between baseline and the final study visit. We will analyze the surrogate marker since it more accurately reflects response adaptive allocation as presented in this paper.

This trial had many design related complications; see \citet{Tamu:Fari:Ande:Heil:ACas:1994} for additional details. In this section the inference procedure accounts for only the permuted block randomization initialization, the RPW(1,1) response-adaptive randomization following the initialization, and the stratified randomization.

For patients in the shortened REML strata the UMLEs were $\hat{p}_{1} = 3/17 = 0.18$ and $\hat{p}_{2} = 7/12 = 0.58$ for the placebo and fluoxetine, respectively. Conversely, the CMLEs were $\tilde{p}_{1} = 0.10$ and $\tilde{p}_{2} = 0.73$. This represents a significantly different picture of the expected probabilities of success from the UMLE. The 95$\%$ simultaneous bootstrap unconditional confidence intervals are (0.00,0.44) for $p_{1}$ and (0.29,0.81) for $p_{2}$. Alternatively, the conditional 95$\%$ simultaneous bootstrap confidence intervals are (0.03, 0.41) for $p_{1}$ and (0.22,0.88) for $p_{2}$.

For patients in the normal REML strata there UMLEs were $\hat{p}_{1} = 10/18 = 0.56$ and $\hat{p}_{2} = 8/14 = 0.57$ for the placebo and fluoxetine, respectively. Conversely, the CMLEs were $\tilde{p}_{1} = 0.53$ and $\tilde{p}_{2} = 0.62$. The 95$\%$ simultaneous bootstrap unconditional confidence intervals are (0.21,0.81) for $p_{1}$ and (0.23,0.82) for $p_{2}$. The conditional 95$\%$ simultaneous bootstrap confidence intervals are (0.27, 0.79) for $p_{1}$ and (0.28,0.84) for $p_{2}$.

Using either conditional or unconditional inference there is little evidence of a fluoxetine benefit with respect to the surrogate marker. However, the conditional intervals were narrower than their unconditional counterpart for 3 of the 4 parameters of interest.

\section{Discussion}

This paper compared conditional and unconditional information and inference in binary response adaptive allocation designs. The primary message is that the intuitive arguments found in \citet{Wei:Smyt:Lin:Stat:1990}, \citet{Begg:OnIn:1990}, \citet{Flem:OnIn:1990}, \citet{Rose:Lach:Rand:2002}, \citet{Anto:Giov:OnTh:2006} and \citet{Pros:Dodd:ReRa:2019}, that the treatment sample sizes contain evidence regarding the effectiveness of the treatments implies conditional information and inference is inferior to unconditional information and inference do not hold in general. Specifically, it was shown that conditional information can be greater than unconditional information,the variance of the CMLE can be less than the variance of the UMLE and conditional confidence intervals can be narrower than relevant unconditional intervals.

To compare information an expression for the conditional information relative to unconditional information was derived. Using relative information it was shown that it is indeed possible for conditional information to exceed unconditional information. Through heuristics it was observed that this is likely to be true even for large samples sizes.

For conditional inference the CMLE was considered in place of the UMLE. The performance of conditional inference was examined in a distribution for three well known response adaptive designs. For each design considered it was found that there exists significant subsets of the parameter space such that the variance of the CMLE was less than the variance of the UMLE. A conditional bootstrap procedure was also proposed that provides consistent coverage. Further, in the majority of cases examined it was found that the this conditional bootstrap procedure resulted in significantly narrower confidence intervals than the relevant unconditional procedures.

A final remark is that conditional inference was not superior to the unconditional inference in every case. Our recommendation to decide which will provide better inference would be to conduct a rigourous simulation study for the design and the expected success probabilities. Then select the inference procedure, conditional or unconditional, that performed best under the expected experimental conditions.

\appendix

\section{Proof of Theorem \ref{thm:CondInfo}}

Denote the likelihood of $\boldsymbol{N}(n)$ as
\begin{align}
L_{\boldsymbol{n}}(\boldsymbol{p};\boldsymbol{n}) = P\{ \boldsymbol{N}(n)=\boldsymbol{n} \}.
\end{align}
Denote the score function of $L_{\boldsymbol{n}}$ as
\begin{align}
\Phi(\boldsymbol{p};\boldsymbol{n}) = \frac{\partial}{\partial \boldsymbol{p}} L_{\boldsymbol{n}}(\boldsymbol{p};\boldsymbol{n}).
\end{align}
The observed information, with respect to $\boldsymbol{p}$, contained in $L_{\boldsymbol{n}}$ is
\begin{align}
\dot{\Phi}(\boldsymbol{p};\boldsymbol{n}) = \frac{\partial}{\partial \boldsymbol{p}} \Phi(\boldsymbol{p};\boldsymbol{n})^{T}.
\end{align}
The following lemma describes the first and second derivatives of $\Phi$.  
\begin{lemma} \label{lem:Partials} Under Conditions \ref{C1}
\begin{align}
\Phi(\boldsymbol{p};\boldsymbol{n}) &= \Lambda_{\boldsymbol{n}}\mbox{E}[\boldsymbol{\hat{p}} - \boldsymbol{p}| \boldsymbol{N}(n)=\boldsymbol{n} ], \\
\dot{\Phi}(\boldsymbol{p};\boldsymbol{n}) &= \Lambda_{\boldsymbol{n}} \mbox{Var}[\boldsymbol{\hat{p}}|\boldsymbol{N}(n)=\boldsymbol{n}] \Lambda_{\boldsymbol{n}} - D(\boldsymbol{p}) \Lambda_{\boldsymbol{n}} - \Lambda_{\boldsymbol{n}}.
\end{align}
\end{lemma}
Proof of Lemma. We use a technique similar to that of \citet{Coad:Ivan:Bias:2001}. For shorthand let $P_{\boldsymbol{p}}$ denote the probability measure of treatment responses and $E_{\boldsymbol{p}}$ denote the expectation with respect to $P_{\boldsymbol{p}}$. Further denote the likelihood ratio
\begin{align}
R(\boldsymbol{p};\boldsymbol{s},\boldsymbol{n})  = \frac{L_{f}(\boldsymbol{p};\boldsymbol{s},\boldsymbol{n})}{L_{f}(\boldsymbol{1/2};\boldsymbol{s},\boldsymbol{n})} = \prod_{k=1}^{K} 2^{n_{k}} p_{k}^{s_{k}}(1-p_{k})^{n_{k}-s_{k}}.
\end{align}
The first and second derivatives of $R(\boldsymbol{p};\boldsymbol{s},\boldsymbol{n})$ can be shown to be
\begin{align}
\dot{R}(\boldsymbol{p};\boldsymbol{s},\boldsymbol{n}) &=  \Lambda_{\boldsymbol{n}} (\boldsymbol{\hat{p}} - \boldsymbol{p}) R(\boldsymbol{p};\boldsymbol{s},\boldsymbol{n})  \\
\ddot{R}(\boldsymbol{p};\boldsymbol{s},\boldsymbol{n}) &= \left\{\dot{\Lambda}_{\boldsymbol{n}}(\boldsymbol{\hat{p}} - \boldsymbol{p}) - \Lambda_{\boldsymbol{n}} + \Lambda_{\boldsymbol{n}} (\boldsymbol{\hat{p}} - \boldsymbol{p}) (\boldsymbol{\hat{p}} - \boldsymbol{p})^{T} \Lambda_{\boldsymbol{n}} \right\} R(\boldsymbol{p};\boldsymbol{s},\boldsymbol{n}) .
\end{align}
The fundamental identity of sequential analysis [see \citet{Wood:Nonl:1982}] ensures that
\begin{align}
\int dP_{\boldsymbol{p}} = \int R(\boldsymbol{p};\boldsymbol{s},\boldsymbol{n})  dP_{\boldsymbol{1/2}},
\end{align}
where $P_{\boldsymbol{1/2}}$ represents the probability measure with $p_{1}=\cdots=p_{K} = 1/2$. Note the assumption that $E_{\boldsymbol{p}}$ is continuous in $\boldsymbol{p}$ ensures derivatives and integrals are interchangeable; therefore,
\begin{align} \label{eq:dbias}
\frac{\partial}{\partial \boldsymbol{p}} P\{ \boldsymbol{N}(n)=\boldsymbol{n} \} &= \frac{\partial}{\partial \boldsymbol{p}} \int I(\boldsymbol{N}(n)=\boldsymbol{n}) dP_{\boldsymbol{p}}  \\
&= \frac{\partial}{\partial \boldsymbol{p}} \int I(\boldsymbol{N}(n)=\boldsymbol{n}) R(\boldsymbol{p};\boldsymbol{s},\boldsymbol{n})  dP_{\boldsymbol{1/2}}   \\
&= \int \Lambda_{\boldsymbol{n}} I(\boldsymbol{N}(n)=\boldsymbol{n}) (\boldsymbol{\hat{p}} - \boldsymbol{p}) R(\boldsymbol{p};\boldsymbol{s},\boldsymbol{n})  dP_{\boldsymbol{1/2}} \\
&=  E_{\boldsymbol{p}}\left[ \Lambda_{\boldsymbol{n}} I(\boldsymbol{N}(n)=\boldsymbol{n}) (\boldsymbol{\hat{p}} - \boldsymbol{p})\right].
\end{align}
From the definition of $\Phi$ we get
\begin{align}
\Phi(\boldsymbol{p};\boldsymbol{n}) &= \frac{\frac{\partial}{\partial \boldsymbol{p}} P\{ \boldsymbol{N}(n)=\boldsymbol{n} \}}{ P\{ \boldsymbol{N}(n)=\boldsymbol{n} \} } \\
&= \Lambda_{\boldsymbol{n}}  \mbox{E}_{\boldsymbol{p}}\left[(\boldsymbol{\hat{p}} - \boldsymbol{p}) | \boldsymbol{N}(n)=\boldsymbol{n} \right]
\end{align}
as stated.

Using steps similar to \eqref{eq:dbias} one can show that
\begin{align}
\frac{\partial^{2}}{\partial \boldsymbol{p}^{2}} P\{ \boldsymbol{N}(n)=\boldsymbol{n} \} &= \mbox{E}_{\boldsymbol{p}}\left[ I(\boldsymbol{N}(n)=\boldsymbol{n}) \{ \dot{\Lambda}_{\boldsymbol{n}}(\boldsymbol{\hat{p}} - \boldsymbol{p}) - \Lambda_{\boldsymbol{n}} + \Lambda_{\boldsymbol{n}} (\boldsymbol{\hat{p}} - \boldsymbol{p}) (\boldsymbol{\hat{p}} - \boldsymbol{p})^{T} \Lambda_{\boldsymbol{n}} \} \right].
\end{align}
Therefore,
\begin{align}
\dot{\Phi}(\boldsymbol{p};\boldsymbol{s},\boldsymbol{n}) &= \frac{\frac{\partial^{2}}{\partial \boldsymbol{p}^{2}} P\{ \boldsymbol{N}(n)=\boldsymbol{n} \} }{P\{ \boldsymbol{N}(n)=\boldsymbol{n} \}}- \frac{\frac{\partial}{\partial \boldsymbol{p}} P\{ \boldsymbol{N}(n)=\boldsymbol{n} \}}{P\{ \boldsymbol{N}(n)=\boldsymbol{n} \}} \left[\frac{\frac{\partial}{\partial \boldsymbol{p}} P\{ \boldsymbol{N}(n)=\boldsymbol{n} \}}{P\{ \boldsymbol{N}(n)=\boldsymbol{n} \}}\right]^{T} \\
&=   \Lambda_{\boldsymbol{n}} \mbox{E}_{\boldsymbol{p}}\left[  (\boldsymbol{\hat{p}} - \boldsymbol{p}) (\boldsymbol{\hat{p}} - \boldsymbol{p})^{T} | \boldsymbol{N}(n)=\boldsymbol{n} \right] \Lambda_{\boldsymbol{n}} + \dot{\Lambda}_{\boldsymbol{n}} \mbox{E}_{\boldsymbol{p}}\left[ (\boldsymbol{\hat{p}} - \boldsymbol{p}) | \boldsymbol{N}(n)=\boldsymbol{n} \right] - \Lambda_{\boldsymbol{n}} \\
&\quad\quad -  \Lambda_{\boldsymbol{n}}  \mbox{E}_{\boldsymbol{p}}\left[(\boldsymbol{\hat{p}} - \boldsymbol{p}) | \boldsymbol{N}(n)=\boldsymbol{n} \right]\mbox{E}_{\boldsymbol{p}}\left[(\boldsymbol{\hat{p}} - \boldsymbol{p}) | \boldsymbol{N}(n)=\boldsymbol{n} \right]^{T}\Lambda_{\boldsymbol{n}} \\
&=\Lambda_{\boldsymbol{n}} \mbox{Var}[\boldsymbol{\hat{p}}|\boldsymbol{N}(n)=\boldsymbol{n}] \Lambda_{\boldsymbol{n}} - D(\boldsymbol{\hat{p}}) \Lambda_{\boldsymbol{n}} - \Lambda_{\boldsymbol{n}}
\end{align}
This concludes the proof of the Lemma.

Returning to the proof of the theorem. Per Lemma \ref{lem:Partials} the conditional score function can be written
\begin{align} \label{eq:CondScore}
\Gamma(\boldsymbol{p};\boldsymbol{s},\boldsymbol{n}) &= \Psi(\boldsymbol{p};\boldsymbol{s},\boldsymbol{n}) - \Phi(\boldsymbol{p};\boldsymbol{n}) \\
&= \Lambda_{\boldsymbol{n}} \left(\boldsymbol{\hat{p}} - \mbox{E}[\boldsymbol{\hat{p}} |\boldsymbol{N}(n)=\boldsymbol{n}] \right).
\end{align}
From the above we find
\begin{align}
\dot{\Gamma}(\boldsymbol{p};\boldsymbol{s},\boldsymbol{n}) &= \Lambda_{\boldsymbol{n}} \mbox{Var}[\boldsymbol{\hat{p}}|\boldsymbol{N}(n)=\boldsymbol{n}] \Lambda_{\boldsymbol{n}} - D(\boldsymbol{\hat{p}}).
\end{align}
Evaluating the above at $\boldsymbol{p} = \boldsymbol{\hat{p}}$ gives the conditional observed information as stated.

From \eqref{eq:CondScore} it is clear that
\begin{align}
\mbox{E}\left[\Gamma(\boldsymbol{p};\boldsymbol{s},\boldsymbol{n}) | \boldsymbol{N}(n)=\boldsymbol{n} \right] &= 0.
\end{align}
Therefore, we can write the conditional expected information
\begin{align}
I_{\boldsymbol{S}|\boldsymbol{N}}(\boldsymbol{\hat{p}}) &= \mbox{E}\left[\boldsymbol{\Gamma}(\boldsymbol{p};\boldsymbol{s},\boldsymbol{n})\boldsymbol{\Gamma}(\boldsymbol{p};\boldsymbol{s},\boldsymbol{n}) ^{T}  | \boldsymbol{N}(n)=\boldsymbol{n}\right] \\
&= \Lambda_{\boldsymbol{n}} \mbox{E}\left[(\boldsymbol{\hat{p}} - \mbox{E}[\boldsymbol{\hat{p}} | \boldsymbol{N}(n)=\boldsymbol{n} ] )(\boldsymbol{\hat{p}} - \mbox{E}[\boldsymbol{\hat{p}}| \boldsymbol{N}(n)=\boldsymbol{n} ])^{T} | \boldsymbol{N}(n)=\boldsymbol{n} \right] \Lambda_{\boldsymbol{n}} \\
&= \Lambda_{\boldsymbol{n}} \mbox{Var}[\boldsymbol{\hat{p}}| \boldsymbol{N}(n)=\boldsymbol{n} ] \Lambda_{\boldsymbol{n}}
\end{align}
as stated.

\section{Proof of Theorem \ref{thm:condMLE}}

From Lemma \ref{lem:Partials} the log-likelihood of the conditional distribution can be written
\begin{align}
\Gamma(\boldsymbol{p};\boldsymbol{s},\boldsymbol{n}) &= \Lambda_{\boldsymbol{n}} \left(\boldsymbol{\hat{p}} - \mbox{E}[\boldsymbol{\hat{p}} | \boldsymbol{N}(n)=\boldsymbol{n}] \right).
\end{align}
Setting equal to 0 and re-arranging terms yields the result of the theorem.

\section{Proof of Theorem \ref{thm:condNorm}}

First, note that E$[\boldsymbol{\hat{p}}|\boldsymbol{N}(n)=\boldsymbol{n}] = \boldsymbol{p} - \boldsymbol{b}_{\boldsymbol{n}}(\boldsymbol{p})$, were $\boldsymbol{b}_{\boldsymbol{n}}(\boldsymbol{p})$ is the bias of $\boldsymbol{\hat{p}}$. Then $\boldsymbol{\tilde{p}}$ satisfies
\begin{align} \label{eq:bias}
\boldsymbol{\tilde{p}} = \boldsymbol{\hat{p}} - \boldsymbol{b}_{\boldsymbol{n}}(\boldsymbol{\tilde{p}}).
\end{align}
By assumption $\boldsymbol{b}_{\boldsymbol{n}}(\boldsymbol{p}) \rightarrow 0$ in probability as $n\rightarrow\infty$; therefore,
\begin{align} \label{eq:convBias}
\boldsymbol{\tilde{p}} - \boldsymbol{p} \rightarrow 0
\end{align}
in probability as $n\rightarrow\infty$. Further from  \eqref{eq:bias}
\begin{align} \label{eq:biasEqual}
(\boldsymbol{\tilde{p}} - \boldsymbol{p}) + \{\boldsymbol{b}_{\boldsymbol{n}}(\boldsymbol{\tilde{p}}) - \boldsymbol{b}_{\boldsymbol{n}}(\boldsymbol{p})\} = \boldsymbol{\hat{p}} - \boldsymbol{p} - \boldsymbol{b}_{\boldsymbol{n}}(\boldsymbol{p})
\end{align}
A Taylor series expansion of the bias with respect to $\boldsymbol{p}$ around $\boldsymbol{p}$ yields
\begin{align} \label{eq:biasExp}
\boldsymbol{b}_{\boldsymbol{n}}(\boldsymbol{\tilde{p}}) = \boldsymbol{b}_{\boldsymbol{n}}(\boldsymbol{p}) + \boldsymbol{\dot{b}}_{\boldsymbol{n}}(\boldsymbol{p})(\boldsymbol{\tilde{p}} - \boldsymbol{p}) + o_{p}(|| \boldsymbol{\tilde{p}} - \boldsymbol{p} ||).
\end{align}
Using the methods of Lemma \eqref{lem:Partials} we show
\begin{align}
\boldsymbol{\dot{b}}_{\boldsymbol{n}}(\boldsymbol{p}) &= \frac{\partial}{\partial \boldsymbol{p}} \mbox{E}[\boldsymbol{\hat{p}} - \boldsymbol{p}|\boldsymbol{N}(n)=\boldsymbol{n}] \\
&= \frac{\partial}{\partial \boldsymbol{p}} \frac{\mbox{E}[\boldsymbol{\hat{p}} - \boldsymbol{p}I(\boldsymbol{N}(n)=\boldsymbol{n})]}{P\{\boldsymbol{N}(n)=\boldsymbol{n}\}} \\
&= \frac{\frac{\partial}{\partial \boldsymbol{p}}\mbox{E}[(\boldsymbol{\hat{p}} - \boldsymbol{p})I(\boldsymbol{N}(n)=\boldsymbol{n})]}{P\{\boldsymbol{N}(n)=\boldsymbol{n}\}} - \frac{\mbox{E}[(\boldsymbol{\hat{p}} - \boldsymbol{p})I(\boldsymbol{N}(n)=\boldsymbol{n})]\frac{\partial}{\partial \boldsymbol{p}}P\{\boldsymbol{N}(n)=\boldsymbol{n}\}}{P\{\boldsymbol{N}(n)=\boldsymbol{n}\}^{2}}.
\end{align}
First, consider
\begin{align}
\frac{\partial}{\partial \boldsymbol{p}}\mbox{E}[(\boldsymbol{\hat{p}} - \boldsymbol{p})I(\boldsymbol{N}(n)=\boldsymbol{n})] &= \int (\boldsymbol{\hat{p}} - \boldsymbol{p})I(\boldsymbol{N}(n)=\boldsymbol{n}) dP_{\boldsymbol{p}} \\
&= \frac{\partial}{\partial \boldsymbol{p}}\int (\boldsymbol{\hat{p}} - \boldsymbol{p})I(\boldsymbol{N}(n)=\boldsymbol{n}) dP_{\boldsymbol{p}} \\
&= \frac{\partial}{\partial \boldsymbol{p}} \int (\boldsymbol{\hat{p}} - \boldsymbol{p})I(\boldsymbol{N}(n)=\boldsymbol{n}) R(\boldsymbol{p};\boldsymbol{s},\boldsymbol{n}) dP_{\boldsymbol{1/2}} \\
&= \int I(\boldsymbol{N}(n)=\boldsymbol{n})[\Lambda_{\boldsymbol{n}} (\boldsymbol{\hat{p}} - \boldsymbol{p})(\boldsymbol{\hat{p}} - \boldsymbol{p})^{T} - \boldsymbol{1}_{K}] R(\boldsymbol{p};\boldsymbol{s},\boldsymbol{n}) dP_{\boldsymbol{1/2}} \\
&= \Lambda_{\boldsymbol{n}}\mbox{E}[(\boldsymbol{\hat{p}} - \boldsymbol{p})(\boldsymbol{\hat{p}} - \boldsymbol{p})^{T}I(\boldsymbol{N}(n)=\boldsymbol{n})] -\boldsymbol{1}_{K}P\{\boldsymbol{N}(n)=\boldsymbol{n}\}. \label{eq:one}
\end{align}
Using the form of $(\partial/\partial \boldsymbol{p})P\{\boldsymbol{N}(n)=\boldsymbol{n}\}$ as given in \eqref{eq:dbias} and \eqref{eq:one} we can rewrite
\begin{align}
\boldsymbol{\dot{b}}_{\boldsymbol{n}}(\boldsymbol{p}) &= \Lambda_{\boldsymbol{n}}\mbox{E}[(\boldsymbol{\hat{p}} - \boldsymbol{p})(\boldsymbol{\hat{p}} - \boldsymbol{p})^{T}|\boldsymbol{N}(n)=\boldsymbol{n}] - \boldsymbol{1}_{K} - \mbox{E}[(\boldsymbol{\hat{p}} - \boldsymbol{p})|\boldsymbol{N}(n)=\boldsymbol{n}]\mbox{E}[(\boldsymbol{\hat{p}} - \boldsymbol{p})|\boldsymbol{N}(n)=\boldsymbol{n}]^{T} \\
&= \Lambda_{\boldsymbol{n}}\mbox{Var}[\boldsymbol{\hat{p}}|\boldsymbol{N}(n)=\boldsymbol{n}] - \boldsymbol{1}_{K}. \label{eq:dbiasVar}
\end{align}
Substituting \eqref{eq:dbiasVar} into \eqref{eq:biasExp}, then further substituting \eqref{eq:biasExp} into \eqref{eq:biasEqual} and re-arranging terms we have
\begin{align} \label{eq:biasAsy}
(\boldsymbol{\tilde{p}} - \boldsymbol{p}) = \{\Lambda_{\boldsymbol{n}} \mbox{Var}[\boldsymbol{\hat{p}}|\boldsymbol{N}(n)=\boldsymbol{n}]\}^{-1} [\boldsymbol{\hat{p}} - \boldsymbol{p} - \boldsymbol{b}_{\boldsymbol{n}}(\boldsymbol{p})] + o_{p}(|| \boldsymbol{\tilde{p}} - \boldsymbol{p} ||).
\end{align}
Equation \eqref{eq:convBias} implies that $o_{p}(|| \boldsymbol{\tilde{p}} - \boldsymbol{p} ||) = o_{p}(1)$. Multiplying both sides of the above by $\left\{I_{\boldsymbol{S}|\boldsymbol{N}}(\boldsymbol{p})\right\}^{1/2}$ as defined in Theorem \ref{thm:CondInfo} we get
\begin{align}
\left\{I_{\boldsymbol{S}|\boldsymbol{N}}(\boldsymbol{p})\right\}^{1/2}(\boldsymbol{\tilde{p}} - \boldsymbol{p})  = \mbox{Var}[\boldsymbol{\hat{p}}|\boldsymbol{N}(n)=\boldsymbol{n}]^{-1/2}\left\{ \boldsymbol{\hat{p}} - \boldsymbol{p} - \boldsymbol{b}_{\boldsymbol{n}}(\boldsymbol{p}) \right\} + o_{p}(1).
\end{align}
Finally, squaring and taking the conditional expectation of both sides of the above yields
\begin{align}
\mbox{E}[\boldsymbol{Z_{\boldsymbol{H}}}^{2}|\boldsymbol{N}(n)=\boldsymbol{n}] = \boldsymbol{I}_{K}[1 + o_{p}(1)],
\end{align}
where $\boldsymbol{H}$ can be either $I_{\boldsymbol{S}|\boldsymbol{N}}(\boldsymbol{p})$ or $J_{\boldsymbol{S}|\boldsymbol{N}}(\boldsymbol{p})$ due to their asymptotic equivalence. Further, since the bias converges to zero in probability as shown by \eqref{eq:convBias} we can replace E$[\boldsymbol{Z_{\boldsymbol{H}}}|\boldsymbol{N}(n)=\boldsymbol{n}]$ with Var$[\boldsymbol{Z_{\boldsymbol{H}}}|\boldsymbol{N}(n)=\boldsymbol{n}]$. Therefore,
\begin{align}
\mbox{Var}[\boldsymbol{Z_{\boldsymbol{H}}}^{2}|\boldsymbol{N}(n)=\boldsymbol{n}] = \boldsymbol{I}_{K}[1 + o_{p}(1)],
\end{align}
which implies the result of the theorem.

\section{Monotonicity of the conditional MLE} \label{app:Mono}
Let $\boldsymbol{h}(\boldsymbol{p}) = \mbox{E}[\boldsymbol{\hat{p}}|\boldsymbol{N}(n)]$ and recall from Theorem \ref{thm:condMLE} that $\boldsymbol{\tilde{p}}$ is defined as the solution to
\begin{align} \label{eq:h}
\boldsymbol{\hat{p}} = \boldsymbol{h}(\boldsymbol{p}).
\end{align}
From \eqref{eq:one} it can be deduced that
\begin{align}
\frac{\partial}{ \partial \boldsymbol{p} } \boldsymbol{h}(\boldsymbol{p}) &= \frac{\partial}{ \partial \boldsymbol{p} } \mbox{E}\left[ \boldsymbol{\hat{p}} | \boldsymbol{N}(n) \right]\\
&= \Lambda_{\boldsymbol{n}} \mbox{Var}[\boldsymbol{\hat{p}}|\boldsymbol{N}(n)=\boldsymbol{n}] \\
&= \mbox{Var}[\boldsymbol{Z}_{n}|\boldsymbol{N}(n)=\boldsymbol{n}] \\
&> 0 \label{eq:pd}.
\end{align}
The assumption that E$_{\boldsymbol{p}}$ is continuous in $\boldsymbol{p}$ ensures that $\boldsymbol{h}$ is continuously differentiable. Therefore, by the inverse function theorem $\boldsymbol{h}$ is invertible near $\boldsymbol{p}$. This implies that if $\boldsymbol{N}(n)$ and $\boldsymbol{S}^{(i)}(n)$ are given then $\boldsymbol{h}$ is a monotone function of $S_{i}(n)$.

\section{Proof of Theorem \ref{thm:boot}}

Let $P^{*}$ represent the empirical probability measure given $\boldsymbol{X}(n) = \{S_{1}(n),\ldots,S_{K}(n),N_{1}(n),\ldots,N_{K-1}(n)\}$. We begin by showing the consistency of the conditional probability measure.
\begin{align}
F^{*}(x) = P^{*}\{\sqrt{n}(\hat{p}_{k}^{*} - \hat{p}_{k}) \le x | \boldsymbol{N}^{*}(n)=\boldsymbol{n} \} &= \frac{ P^{*}\{ [\sqrt{n}(\hat{p}_{k}^{*} - \hat{p}_{k}) \le x] \bigcap [\boldsymbol{N}^{*}(n)=\boldsymbol{n}] \}}{ P^{*}\{ \boldsymbol{N}^{*}(n)=\boldsymbol{n} \} } \\
&= \frac{ P\{ [\sqrt{n}(\hat{p}_{k} - p_{k}) \le x] \bigcap [\boldsymbol{N}(n)=\boldsymbol{n}] \} + o_{p}(1)}{ P\{ \boldsymbol{N}(n)=\boldsymbol{n} \} + o_{p}(1) } \\
&= P\{\sqrt{n}(\hat{p}_{k} - p_{k}) \le x | \boldsymbol{N}(n)=\boldsymbol{n} \} + o_{p}(1).
\end{align}
The above ensures the consistency of the inverse of $F^{*}$ at $\alpha$, i.e.,
\begin{align}
F^{*-1}(\alpha) \rightarrow F^{-1}(\alpha)
\end{align}
in probability as $n\rightarrow\infty$. Let $h_{k}$ be the $k$th element of $\boldsymbol{h}=(h_{1},\ldots,h_{K})^{T}$ as defined in \eqref{eq:h}, then by continuity and monotonicity we have that
\begin{align}
\tilde{p}_{k}^{\alpha} = h_{k}^{-1}\left\{F^{*-1}(\alpha)\right\} \rightarrow h_{k}^{-1}\left\{F^{-1}(\alpha)\right\}.
\end{align}
in probability as $n\rightarrow\infty$. The right hand side of the above is the solution with respect to $x$ of
\begin{align}
P\{\sqrt{n}(\tilde{p}_{k} - p_{k}) \le x | \boldsymbol{N}(n)=\boldsymbol{n} \} = \alpha.
\end{align}
This completes the proof.

\bibliographystyle{elsarticle-harv}
\bibliography{bibtex_entries}

\end{document}